\documentclass[graybox]{svmult}

\usepackage[utf8]{inputenc}

\usepackage{type1cm}
\usepackage{newtxtext}
\usepackage{newtxmath}

\usepackage{makeidx}
\usepackage{multicol}
\usepackage[bottom]{footmisc}

\usepackage{amsmath}
\usepackage{hyperref}
\usepackage{graphicx}
\usepackage{color}
\usepackage{bm}

\DeclareMathOperator{\diag}{\mathrm{diag}}
\renewcommand{\Re}{\operatorname{Re}}

\makeindex

\begin{document}

\title*{Symmetry Approach to Chiral Optomagnonics in Antiferromagnetic Insulators}
\author{Igor Proskurin and Robert L. Stamps} 
\institute{
  Igor Proskurin \at Department of Physics \& Astronomy, University
  of Manitoba, Winnipeg, MB R3T~2N2 Canada; Institute of Natural
  Sciences and Mathematics, Ural Federal University, Ekaterinburg,
  620002 Russia; \email{Igor.Proskurin@umanitoba.ca} \and Robert
  L. Stamps \at Department of Physics \& Astronomy, University of
  Manitoba, Winnipeg, MB R3T~2N2 Canada;
  \email{Robert.Stamps@umanitoba.ca} }

\maketitle

\abstract{ We discuss several aspects of chiral optomagnonics in
  antiferromagnetic insulators by considering common symmetries
  between the electromagnetic field and spin excitations. This approach
  allows us to look at optical and magnetic materials from similar
  perspectives, and discuss useful analogies between them. We show
  that spin waves in collinear antiferromagnets and the
  electromagnetic field in vacuum are both invariant under the same
  eight-dimensional algebra of symmetry transformations. By
  such analogy, we can extend the concept of optical chirality to
  antiferromagnetic insulators, and demonstrate that the spin-wave
  dynamics in these materials in the presence of a spin current is 
  similar to that of the light inside chiral
  metamaterials. Photo-excitation of magnonic spin currents is also
  discussed from the symmetry point of view. It is demonstrated that a
  direct magnonic spin photocurrent can be exited by circularly
  polarized light, which can be considered as a magnonic analogue of the
  photogalvanic effect.  We also note that the \emph{Zitterbewegung} process should appear and may play a role in photo-excitation processes.}

\section{Introduction}
\label{sec:1}
Modern spintronics is now a well-developed area that aims at bringing new functionality to
conventional electronics by making use of the spin degrees of freedom
\cite{Zutic2004}, which may help to overcome looming saturation of Moore's Law \cite{Thomson2006}. There are a number of different
trends in the development of the spintronics today. Among different
materials, antiferromagnets play an important role, which brings us to
the field of antiferromagnetic spintronics \cite{Baltz2018,
  Smejkal2018}. Their abundance in Nature and zero net magnetization
make antiferromagnets potentially useful for applications, while the existence
of two or more magnetic sublattices allows one to explore various
topological effects \cite{Smejkal2018}. The focus on optical
manipulation of the spin states in magnetic insulators constitutes the
scope of the optospintronics \cite{Nemec2018}. A prominent direction
in optospintronics is related to the application of microwave
cavity resonators \cite{Harder2018a}, which has already seen a rapid
development during the last several years \cite{Kusminskiy2019}.

Being interdisciplinary, spintronics in general, and optomagnonics in
particular, can benefit by looking at the concept of
chirality. Chirality or handedness, which according to the original
definition given by Lord Kelvin in his Baltimore Lectures is related
to the lack of symmetry between an object and its mirror image
\cite{Kelvin1904}.  It is a universal phenomenon that has proved its
significance in various scientific areas from high-energy physics to life sciences and soft matter
\cite{Barron2012}. Kelvin's definition, which is purely geometric, was
generalized later to accommodate dynamical phenomena
by Barron \cite{Barron1986}. Thus, according to Barron's definition, one should
distinguish between \emph{true} and \emph{false} chiralities. The
former is to be found in the systems that break inversion
symmetry, but at the same time are invariant under a time-reversal
transformation combined with any proper rotation, while the latter is
characterized by breaking time-reversal and inversion symmetries
simultaneously \cite{Barron2004}.

How can the concept of chirality be useful for the development of
optospintronics? A general observation is that the goal of the
spintronics is manipulation and transformation of pure spin currents,
and spin currents are chiral. Indeed, in agreement with the definition
of true chirality, a flow of angular momentum reverses its sign under spatial inversion, while it remains invariant
under the time reversal transformation, which reverses both velocities and spins. Thus, from the symmetry point
of view, pure spin currents are in the same category as, for example,
natural optical activity and circular dichroism in optics. This
argument also suggests that materials with structural chirality may have unique properties for hosting and transferring spin currents that makes them interesting for applications, which is reflected in the rapid development of
molecular spintronics \cite{Naaman2015, Michaeli2016} and related
topics such as chiral spin selectivity \cite{Naaman2019}.

Another observation helpful to establish a link between optics and
spintronics is that not only geometric structures but also physical fields can be characterized by chirality. Chirality density of the electromagnetic field, for example, has
been known for a long time. Lipkin first noticed that the Maxwell's
equations in vacuum have a hidden conservation
law for a chiral density, which he dubbed \emph{zilch} due to the lack
of clear physical meaning of this quantity at that time \cite{Lipkin1964}. Later, it
was demonstrated that this conservation law is closely related to
electromagnetic duality \cite{Calkin1965, Zwanziger1968}. This
eventually led to the formulation of the \emph{nongeometric}
symmetries of the Maxwell's equations \cite{Fushchich1987}, i.~e. the
symmetries, which are not reduced to space-time
transformations. For several decades, the formal properties of optical
chirality, helicity, and dual symmetries were discussed
\cite{Krivskii1989, Krivskii1989a, Afanasiev1996,
  Bialynicki-Birula1996, Drummond1999, Drummond2006, Ibragimov2008,
  Berry2009} but it was not until Tang and Cohen showed how electromagnetic
chirality density can be used to characterize dichroism in light
interacting with a chiral metamaterial that this was understood for materials \cite{Tang2010}. This revived
interest in optical chirality \cite{Bliokh2011, Barnett2012,
  Coles2012, Philbin2013}, which has found a number of applications in
optics and plasmonics \cite{Hendry2010, Tang2011,
  Schaeferling2012, Hendry2012, Kamenetskii2013,
  Canaguier-Durand2013}.

The results of Tang and Cohen \cite{Tang2010} can be understood
as follows. In order to observe effects related to the chirality
of light, we have to put the electromagnetic field in contact with a chiral
environment. This principle suggests a way for finding similar effects
in other systems. For example, spin-wave dynamics in collinear
antiferromagnets can be represented in a form that closely resembles
the Silberstein-Bateman formulation of the Maxwell's equations. Since
collinear antiferromagnets have two magnetic sublattices, the concept
of electromagnetic duality and nongeometric symmetries can be
generalized to transformations between the antiferromagnetic sublattices
\cite{Proskurin2017b}. This allows to establish a conservation law for
a spin-wave analogue of the optical chirality. Injection of a spin
current into the antiferromagnet in this case has an effect similar to
a chiral environment for light-matter interactions inside a
metamaterial \cite{Proskurin2017b}.

It is also remarkable that both the Maxwell's equations
\cite{Barnett2014} and the dynamics of antiferromagnetic spin waves
\cite{Wang2017} allow a formulation in the form of the Dirac equation
for an ultra-relativistic particle. Such particles are
characterized by conserving helicity --- a projection of spin on the linear momentum \cite{Landau1983}, which also satisfies the definition of true chirality. Breaking the symmetry between right and left, in
this case, corresponds to a Weyl material \cite{Yan2017}, wherein
quasi-particles with different helicities are spatially
separated. Symmetry considerations suggest that as far as single
particle dynamics is concerned, there should be some analogy between optical
metamaterials, Weyl semimetals, and chiral antiferromagnets. There has been several proposals in these directions. For example, one can emulate the chiral magnetic effect in metallic antiferromagnets \cite{Sekine2016}.

These arguments have a direct impact on optospintronics. Since optical
chirality and spin currents share the same symmetry properties, it is
possible to use polarized light to excite magnon spin-photocurrents
in antiferromagnetic insulators \cite{Proskurin2018a}. Circular
polarized light in this case creates a direct \emph{flow} of magnon angular
momentum, whose direction is controlled by helicity of light. This
effect resembles the circular photogalvanic effect in metals
\cite{Belinicher1980}, which recently attracted attention in
topological electron materials \cite{Juan2017}. It has been
demonstrated that for a separated Weyl node, the photocurrent
excitation rate is determined by the product of the topological charge of the node
and the helicity of light \cite{Juan2017}.

In this Chapter, we review chiral excitations in optics and antiferromagnetic
insulators together with their applications in optomagnonics. Our discussion is
organized as follows. In Section~\ref{sec:2}, we give a brief review
of optical chirality and nongeometric symmetries, which is
generalized to antiferromagnetic spin-waves in Section~\ref{sec:3}, where we discuss potential applications such as spin-current
induced magnon dichroism. Section~\ref{sec:4} is reserved for photo-excitation of magnon spin currents with polarized light. 
Summary and conclusions are in Section~\ref{sec:5}.

\section{Optical chirality and nongeometric symmetries of the Maxwell's equations}
\label{sec:2}
Since the early developments of electrodynamics, it has been well
established that the electromagnetic field in vacuum can be
characterized by conserving energy, momentum, angular momentum, which reflects the invariance of the Maxwell's
equations with respect to the translations and rotations in the
four-dimensional space-time \cite{Fushchich1987}. It was found almost by chance
\cite{Lipkin1964} that in addition to these conservation laws, the
electromagnetic field has another invariant given by a combination of
the electric, $\bm{E}$, and magnetic, $\bm{B}$, fields
\begin{equation}
  \label{zil}
  \rho_{\chi}(t,\bm{r}) = \frac{\varepsilon_{0}}{2} \bm{E}\cdot(\bm{\nabla}\times\bm{E})
  + \frac{1}{2\mu_{0}}\bm{B}\cdot(\bm{\nabla}\times\bm{B}),
\end{equation}
which is odd under the spatial inversion and even under the time
reversal transformations ($\varepsilon_{0}$ and $\mu_{0}$ are the vacuum permittivity and permeability respectively). For this quantity, Lipkin coined a special term --- optical \emph{zilch} to emphasize the lack of a clear physical
interpretation at that time \cite{Lipkin1964}. According to its
symmetry properties, $\rho_{\chi}$ is \emph{truly chiral}
\cite{Barron1986}, and can be considered as a chirality density of the
electromagnetic field.

Using the Maxwell's equations, it is straightforward to demonstrate
that in vacuum $\rho_{\chi}$ satisfies the continuity equation
\begin{equation}
  \frac{\partial \rho_{\chi}}{\partial t} + \bm{\nabla} \cdot \bm{J}_{\chi} = 0,
\end{equation}
where
\begin{equation} \label{jchi} \bm{J}_{\chi}(t,\bm{r}) =
  \frac{\varepsilon_{0}}{2} \bm{E} \times \frac{\partial
    \bm{E}}{\partial t} + \frac{1}{2\mu_{0}} \bm{B} \times
  \frac{\partial \bm{B}}{\partial t},
\end{equation}
determines the corresponding zilch flow.

In this section, we will show that this conservation law belongs to
the class of so-called ``hidden'' or \emph{nongeometric} symmetries of
the Maxwell's equations. One of these symmetries, which has been known
since the time of Heaviside, Larmor, and Rainich, is the duality
symmetry \cite{Cameron2012, Bliokh2013}. If we consider Maxwell's
equations in free space
\begin{eqnarray}
  \bm{\nabla} \times \bm{E} &=& 0, \qquad  \bm{\nabla} \times \bm{B} = 0, \label{Max1} \\
  \bm{\nabla} \cdot \bm{E}  &=& 0,  \qquad  \bm{\nabla} \cdot \bm{B} = 0, \label{Max2}
\end{eqnarray}
(we set $c = 1$ throughout this section) the electromagnetic duality is
a symmetry with respect to the rotation in the pseudo-space of the
electric and magnetic fields, which leaves Maxwell's equations
invariant
\begin{eqnarray}
  \bm{E} \to \bm{E}' &=& \bm{E} \cos\theta + \bm{B}\sin\theta, \label{du1} \\
  \bm{B} \to \bm{B}' &=& -\bm{E} \sin\theta + \bm{B}\cos\theta,\label{du2}
\end{eqnarray}
where $\theta$ is a real parameter of the transformation. This
symmetry is usually broken inside materials, unless we deal with a
dual symmetric medium \cite{Fernandez-Corbaton2013}.

The existence of duality symmetry guarantees the conservation of optical helicity, i.~e. the projection of spin angular
momentum of the photon onto its linear momentum \cite{Calkin1965,
  Zwanziger1968, Drummond1999, Drummond2006}. It should be mentioned,
however, that the formulation of helicity conservation law in
classical electrodynamics is not straightforward, because the standard
Lagrangian for the electromagnetic field is not dual symmetric
\cite{Bliokh2013}. Using the dual symmetric representation for the
electromagnetic Lagrangian combined with the Noether's approach, it is
possible to express the optical helicity density in the form similar to
Eq.~(\ref{zil})
\begin{equation}
  \rho_{\mathrm{hel}}(t,\bm{r}) = \frac{1}{2}\left[\bm{A}\cdot(\bm{\nabla} \times \bm{A})
    + \bm{C}\cdot(\bm{\nabla} \times \bm{C})  \right],
\end{equation}
where in addition to the magnetic vector potential $\bm{A}$, we also
introduced the electric vector potential $\bm{C}$, which satisfies the
following equations,
$\bm{E} = -\bm{\nabla} \times \bm{C} = -\partial_{t}\bm{A}$ and
$\bm{B} = \bm{\nabla} \times \bm{A} = \partial_{t}\bm{C}$. These are
invariant under the transformations in Eqs.~(\ref{du1}) and
(\ref{du2}) \cite{Cameron2012, Bliokh2013}.

The definition of electromagnetic helicity depends
on a specific representation of the Lagrangian. It suggests that it would
be useful to have a general formalism for deriving ``hidden''
symmetries and conservation laws directly from the equations of motion
formulated exclusively in terms of the electromagnetic fields, and
independent of any gauge choice. Such a formalism has been developed
by Fushchich and Nikitin \cite{Fushchich1987}. Below, we give a brief
review of this formalism, which is necessary for further discussions.

\subsection{Symmetry analysis of the Maxwell's equations}
For the symmetry analysis, it is convenient to formulate Maxwell's
equations in the form that resembles the Dirac equation for a massless
relativistic particle. This representation is called the
Silberstein-Bateman form \cite{Fushchich1987}. In this form, the first
pair of the Maxwell's equations in Eq.~(\ref{Max1}) is rewritten in
terms of a Schr\"{o}dinger-like equation for the six-component vector column
composed of the components of the electric and magnetic fields
$\phi = (\bm{E},\bm{B})^{T}$
\begin{equation} \label{SBM} i\frac{\partial \phi(t,\bm{p})}{\partial
    t} = \mathcal{H}(\bm{p}) \phi(t,\bm{p}),
\end{equation}
where for convenience, we work in the momentum space, $\bm{p}$, defined by the
following Fourier transformations
\begin{eqnarray}
  \bm{E}(t,\bm{r}) &=& \frac{1}{(2\pi)^{3/2}} \int d^{3}p e^{i\bm{p}\cdot\bm{r}} \bm{E}(t,\bm{p}), \\
  \bm{B}(t,\bm{r}) &=& \frac{1}{(2\pi)^{3/2}} \int d^{3}p e^{i\bm{p}\cdot\bm{r}} \bm{B}(t,\bm{p}).
\end{eqnarray}
The matrix on the right-hand side of Eq.~(\ref{SBM}) has the following
structure
\begin{equation}
  \label{eq:Hem}
  \mathcal{H}(\bm{p}) = \left(
    \begin{array}{cc}
      0 & i(\hat{\bm{S}}\cdot\bm{p}) \\
      -i(\hat{\bm{S}}\cdot\bm{p}) & 0
    \end{array}
  \right),
\end{equation}
which can be considered as a direct product of the Pauli matrix
$\sigma_{2}$, which interchanges $\bm{E}$ and $\bm{B}$, and the
``helicity'' operator $(\hat{\bm{S}} \cdot \bm{p})$, where the
matrices $\hat{S}_{\alpha}$ ($\alpha = x, y, z$) form a
representation of the three-dimensional rotation group,
$(\hat{S}_{\alpha})_{\beta\gamma} = i\epsilon_{\alpha\beta\gamma}$, with $\epsilon_{\alpha\beta\gamma}$ being the Levi-Civita symbol.

The second pair of the Maxwell's equations (\ref{Max2}) in this
formalism impose an additional constraint on the components of
$\phi(t, \bm{p})$ \cite{Fushchich1987}
\begin{equation} \label{L2} (\hat{\bm{S}} \cdot \bm{p})^{2}
  \phi(t,\bm{p}) = p^{2}\phi(t,\bm{p}),
\end{equation}
which acknowledges transversality of the electromagnetic field in vacuum.

\subsubsection{Invariance algebra of the Maxwell's equations}
\label{sec:2.1.1}
Now, we can find the symmetry operations that transform a
solution $\phi(t,\bm{p})$ of Eq.~(\ref{SBM}) into another solution
$\tilde{\phi}(t,\bm{p}) = \mathcal{Q}(\bm{p})\phi(t,\bm{p})$. We look
for these transformations in the form of the six-dimensional matrices
$\mathcal{Q}(\bm{p})$, which may depend on the momentum $\bm{p}$.
Formal resemblance of our representation with the quantum mechanics
implies that these matrices should commute with
$\mathcal{H}(\bm{p})$.

The problem of finding all such transformation becomes almost trivial
if we transform to the helicity basis, where $\mathcal{H}(\bm{p})$ is
diagonal. This transformation is reached by a combination of the
rotation in the three-dimensional space
\begin{equation} \label{UL} \hat{U}_{\Lambda} = \left(
    \begin{array}{ccc}
      -\dfrac{p_{x}p_{z} + ip_{y}p}{\sqrt{2}pp_{\perp}} & \dfrac{p_{x}p_{z} - ip_{y}p}{\sqrt{2}pp_{\perp}} & \dfrac{p_{x}}{p} \\
      -\dfrac{p_{y}p_{z} - ip_{x}p}{\sqrt{2}pp_{\perp}} & \dfrac{p_{y}p_{z} + ip_{x}p}{\sqrt{2}pp_{\perp}} & \dfrac{p_{y}}{p} \\
      \dfrac{p_{\perp}}{\sqrt{2}p} & -\dfrac{p_{\perp}}{\sqrt{2}p} & \dfrac{p_{z}}{p}
    \end{array}
  \right),
\end{equation}
where $p_{\perp} = (p_{x}^{2} + p_{y}^{2})^{1/2}$, which diagonalizes
the ``helicity'' operator,
$\hat{U}_{\Lambda}^{\dag}(\hat{\bm{S}} \cdot \bm{p})\hat{U}_{\Lambda} = \diag(-p,p,0)$,
with the $SU(2)$ transformation in the pseudo-space of electric
and magnetic fields
\begin{equation} \label{U2} U_{2} = \frac{1}{\sqrt{2}} \left(
    \begin{array}{cc}
      1 & -i \\
      -i & 1
    \end{array}
  \right).
\end{equation}
The resulting transformation
$\mathcal{U} = U_{2}\otimes\hat{U}_{\Lambda}$ diagonalizes
$\mathcal{H}(\bm{p})$ so that in the transformed frame
\begin{equation} \label{tH} \tilde{\mathcal{H}} = \mathcal{U}^{\dag}
  \mathcal{H} \mathcal{U} = \diag(-p,p,0,p,-p,0).
\end{equation}
The eigenvalues of $\tilde{\mathcal{H}}$ correspond to the left and
right polarized electromagnetic modes with the linear frequency
dispersion $cp$ (we have recovered the speed of light $c$ here), which are degenerate in the absence of light-matter
interactions.

Straightforward calculations show that in the diagonal frame, any
matrix that commutes with $\tilde{\mathcal{H}}$, and at the same time
leaves Eq.~(\ref{L2}) invariant, is parameterized by eight parameters,
$a$, \ldots $h$, and has the following structure
\begin{equation}
  \label{eq:Q1}
  \tilde{\mathcal{Q}} = \left(
    \begin{array}{cccccc}
      a &   0    & 0 &   0    & e & 0 \\
      0    & b & 0 & f &   0    & 0 \\
      0    &   0    & 0 &   0    &   0    & 0 \\
      0    & g & 0 & c &   0    & 0 \\
      h &   0    & 0 &   0    & d & 0 \\
      0    &   0    & 0 &   0    &   0    & 0
    \end{array}
  \right).
\end{equation}
The basis in the linear space of $\tilde{\mathcal{Q}}$ can be chosen
such as its basis elements, $\tilde{\mathcal{Q}}_{i}$,
($i = 1, \dots, 8$) form the algebra isomorphic to the Lie algebra of
the group $U(2) \otimes U(2)$
\begin{equation}
  \begin{array}{rclrcl}
  \tilde{\mathcal{Q}}_{1} &=& -\sigma_{2} \otimes \hat{S}_{y}, \qquad&
  \tilde{\mathcal{Q}}_{2} &=&  -i\sigma_{3} \otimes \hat{I} \\
  \tilde{\mathcal{Q}}_{3} &=& -i\sigma_{1} \otimes \hat{S}_{y}, \qquad&
  \tilde{\mathcal{Q}}_{4} &=&    \sigma_{1} \otimes \hat{S}_{x} \\
  \tilde{\mathcal{Q}}_{5} &=&  -\sigma_{0} \otimes \hat{S}_{z}, \qquad&
  \tilde{\mathcal{Q}}_{6} &=&   \sigma_{2} \otimes \hat{S}_{x} \\
  \tilde{\mathcal{Q}}_{7} &=&  \sigma_{0} \otimes \hat{I}, \qquad&
  \tilde{\mathcal{Q}}_{8} &=&  i\sigma_{3} \otimes \hat{S}_{z},
  \end{array}
\end{equation}
where $\sigma_{0}$ and $\hat{I}$ denote $2 \times 2$ and $3 \times 3$ unit matrices respectively.

Returning into original frame and taking into account that
$\hat{U}_{\Lambda}\hat{S}_{z}\hat{U}_{\Lambda}^{\dag}=-(\hat{\bm{S}}
\cdot \bm{p})/p$, we obtain the generators of the symmetry
transformations in the following form
\begin{equation}
  \label{eq:Q}
  \begin{array}{rclrcl}
  \mathcal{Q}_{1} &=& \sigma_{3}\otimes (\hat{\bm{S}} \cdot \tilde{\bm{p}})\hat{D}, \qquad &
  \mathcal{Q}_{2} &=& i\sigma_{2}\otimes \hat{I}, \\
  \mathcal{Q}_{3} &=& -\sigma_{1}\otimes (\hat{\bm{S}} \cdot \tilde{\bm{p}})\hat{D},\qquad &
  \mathcal{Q}_{4} &=& -\sigma_{1}\otimes \hat{D}, \\
  \mathcal{Q}_{5} &=& \sigma_{0}\otimes (\hat{\bm{S}} \cdot \tilde{\bm{p}}), \qquad &
  \mathcal{Q}_{6} &=& -\sigma_{3}\otimes \hat{D}, \\
  \mathcal{Q}_{7} &=&  \sigma_{0}\otimes \hat{I}, \qquad &
  \mathcal{Q}_{8} &=& i\sigma_{2}\otimes (\hat{\bm{S}} \cdot \tilde{\bm{p}}), 
  \end{array}
\end{equation}
where $\tilde{\bm{p}} = \bm{p}/p$, and
$\hat{D} =
-p\hat{U}_{\Lambda}\hat{S}_{x}\hat{U}_{\Lambda}^{\dag}$. These
equations form the eight-dimensional invariance algebra of the
Maxwell's equations in vacuum \cite{Fushchich1987}.

\subsubsection{Nongeometric symmetries}
The basis elements in Eqs.~(\ref{eq:Q}) generate continuous
symmetries that Fushchich and Nikitin called the \emph{nongeometric
  symmetries} of the Maxwell's equations \cite{Fushchich1987}
\begin{equation}\label{sym}
  \phi(t,\bm{p}) \to \phi'(t,\bm{p}) = \exp(\mathcal{Q}_{i}\theta_{i})\phi(t,\bm{p}),
\end{equation}
where $\theta_{i}$ denotes the real parameter of the transformation.

Some symmetry generators have a clear physical meaning. For
example, $\mathcal{Q}_{7}$ is a unit element. $\mathcal{Q}_{2}$
interchanges electric and magnetic fields in
$\phi(t,\bm{p})$, so that the corresponding continuous transformation
$\exp(i\sigma_{2}\theta)$ is the duality symmetry in Eq.~(\ref{du1})
and (\ref{du2}).  $\mathcal{Q}_{5}$ has the form of the helicity
operator. $\mathcal{Q}_{8}$ is proportional to $\mathcal{H}$, which
means that similar to $\mathcal{Q}_{7}$ it commutes with every
element of the algebra. It reflects the symmetry with respect to
$\partial_{t}$ (the time derivative of $\phi(t,\bm{p})$, which solves the Maxwell's equations, is again a solution for the same
$\bm{p}$). The basis elements $Q_{2}$, $Q_{5}$, $Q_{7}$, and $Q_{8}$ form a
trivial Abelian part of the algebra in
Eqs.~(\ref{eq:Q}). The existence of non-Abelian elements is
related to the degeneracy between left and right polarized
eigenvalues in Eq.~(\ref{tH}).

The conservation laws that correspond to the symmetry transformations
in Eq.~(\ref{sym}) can be conveniently written in terms of the
bilinear forms by analogy with the quantum-mechanics
\begin{equation}
  \label{eq:cons}
  \langle Q_{i} \rangle = \frac{1}{2} \int d^{3}p \phi^{\dag}(t,\bm{p}) Q_{i} \phi(t,\bm{p}).
\end{equation}
It can be demonstrated that the electromagnetic field in vacuum can be
characterized by an infinite number of invariants generated from the
eight symmetry transformations \cite{Fushchich1987}.  For example, the
unit element $\mathcal{Q}_{7}$ in this formalism corresponds to the
conservation of the electromagnetic energy
\begin{equation}
  \langle Q_{7} \rangle = \frac{1}{2} \int d^{3}p \phi^{\dag}(t,\bm{p}) \phi(t,\bm{p}) = \frac{1}{2} \int d^{3}p \left( E^{2} + B^{2} \right). 
\end{equation}

\subsubsection{Conservation law for optical chirality}
\label{sec:2.1.3}
Using this formalism, optical zilch can be expressed as a
conservation law for the helicity operator $\mathcal{Q}_{5}$
\begin{equation} \label{Cchi} C_{\chi} = \int d^{3}r
  \rho_{\chi}(t,\bm{r}) = \frac{1}{2} \int d^{3}p
  \phi^{\dag}(t,\bm{p}) (\hat{\bm{S}} \cdot \bm{p}) \phi(t,\bm{p}).
\end{equation}
Using the fact that the helicity operator, duality symmetry, and
$\partial_{t}$ are related to each other by the algebraic property,
$p\mathcal{Q}_{5}\mathcal{Q}_{2} = -i\mathcal{H} = \partial_{t}$, we
establish a relation between zilch conservation and duality symmetry as it was originally discussed in
\cite{Calkin1965,Zwanziger1968}, which allows to write the
conservation law above in the following equivalent form
\begin{equation}
  C_{\chi} = -\frac{i}{2} \int d^{3}p \phi^{\dag}(t,\bm{p})\mathcal{Q}_{2}\partial_{t}\phi(t,\bm{p})=
  \frac{1}{2}\int d^{3}r \left( \bm{B} \cdot \frac{\partial \bm{E}}{\partial t} - \bm{E} \cdot \frac{\partial \bm{B}}{\partial t} \right).
\end{equation}
This expression can be easily
generalized to accommodate higher order terms in space and time derivatives. By replacing
$Q_{2}\partial_{t}$ with
$-(ip)^{2n}\mathcal{Q}_{2}(i\partial_{t})^{2m+1}$, which is again a
symmetry operation, we can find a hierarchy of conserving zilches
\begin{equation} \label{Cmn} C_{\chi}^{(m,n)} = \frac{1}{2}\int d^{3}r
  \left( \bm{B}\cdot\nabla^{2n}\partial^{2m+1}_{t}\bm{E} -
    \bm{E}\cdot\nabla^{2n}\partial^{2m+1}_{t}\bm{B} \right),
\end{equation}
where $00$-zilch corresponds to the optical chirality
\cite{Drummond1999,Drummond2006, Philbin2013}.

It is possible to derive the conservation law for the optical zilch using the
Noether's formalism by applying a specific ``hidden'' gauge
transformation to the Lagrangian of the electromagnetic field
\cite{Philbin2013}, which leads to the same results as in
Eqs.~(\ref{Cchi}) and (\ref{Cmn}). The advantage of the approach
discussed in this section, based on the symmetry analysis of the
Maxwell's equations, is that it does not depend on any specific gauge
choice. This fact makes it easy to extend this formalism to other
physical systems with similar form of the equations of motion.

\subsection{Optical chirality in gyrotropic media}
\label{sec:2.2}
Having now a complete picture of the nongeometric symmetries in
vacuum, we discuss how this approach can be applied for the
light-matter interactions.  Electromagnetic field in dielectric
medium is usually described by the material form of the Maxwell
equations
\begin{eqnarray}
  \bm{\nabla} \times \bm{E} = -\frac{\partial \bm{B}}{\partial t}, \qquad \bm{\nabla}
  \cdot \bm{B} = 0, \label{Max3} \\
  \bm{\nabla} \times \bm{H} =  \frac{\partial \bm{D}}{\partial t}, \qquad \bm{\nabla} \cdot \bm{D} = 0, \label{Max4}
\end{eqnarray}
supplemented by the constituent relations between the fields $\bm{E}$,
$\bm{H}$, $\bm{D}$, and $\bm{B}$. The constituent relations impose
additional constraints on the form of the symmetry transformations for
the electromagnetic field, which reflect the intrinsic symmetries of
the medium. This often leads to the reduction of the invariance
algebra in Eqs.~(\ref{eq:Q}) to lesser number of elements
\cite{Proskurin2017}.

As an important example, let us consider propagation of the
electromagnetic field in chiral media where structural
chirality of the material leads to the existence of such physical
phenomena as natural optical activity and circular dichroism. There
exists several approaches for the electrodynamics of chiral
gyrotropic media \cite{Fedorov1976, Lekner1996, Cho2015}. One of these
approaches, which is frequently adopted for characterizing
metamaterials \cite{Jaggard1979, Tomita2014}, is based on the
following constituent relations
\begin{eqnarray}
  \bm{D} &=& \varepsilon \varepsilon_{0} \bm{E} + i \varkappa \bm{H}, \label{con1} \\
  \bm{B} &=& \mu \mu_{0} \bm{H} - i \varkappa \bm{E}, \label{con2}
\end{eqnarray}
where $\varepsilon$ and $\mu$ are the electric permittivity and magnetic permeability of the medium respectively, and $\varkappa$ characterizes chirality of the material. This
approach requires complex representation for the electromagnetic
fields and can be derived from the relativistic covariance principle
\cite{Fedorov1976, Post1962}.

By applying our general formalism to the Maxwell's equations
(\ref{Max3}) and (\ref{Max4}) with the constituent relations
(\ref{con1}) and (\ref{con2}), we obtain the same equation of motion
as in Eq.~(\ref{SBM}), where $\phi$ is replaced by for the vector
column $\phi(t,\bm{p}) = (\bm{D}, \bm{B})^{T}$, and the matrix on the
right-hand side is now given by (we use the units
$\varepsilon\varepsilon_{0} = \mu\mu_{0} =1$)
\begin{equation}
  \mathcal{H}(\bm{p}) = -\frac{1}{1-\varkappa^{2}} \left(
    \begin{array}{cc}
      \varkappa (\hat{\bm{S}} \cdot \bm{p}) & -i(\hat{\bm{S}} \cdot \bm{p}) \\
      i(\hat{\bm{S}} \cdot \bm{p}) & \varkappa (\hat{\bm{S}} \cdot \bm{p})
    \end{array}
  \right).
\end{equation}
This matrix can be diagonalized by a combination of the same unitary
transformations as in Eqs.~(\ref{UL}) and (\ref{U2}) that yields the
following diagonal form
\begin{equation}\label{Hcme}
  \tilde{\mathcal{H}} = \mathcal{U}^{\dag} \mathcal{H} \mathcal{U} =
  \diag(-p_{-},p_{-},0, p_{+},-p_{+},0),
\end{equation}
where $p_{\pm} = p/(1 \mp \varkappa)$.

Lifted degeneracy between left ($p_{-}$) and right ($p_{+}$) polarized
eigenmodes in Eq.~(\ref{Hcme}) leads to the reduction of the
eight-dimensional invariance algebra to four basis elements, which
commute with each other
\begin{equation}
  \begin{array}{lcrlcr}
    \mathcal{Q}_{2} &=& i\sigma_{2}\otimes \hat{I}, \qquad & \mathcal{Q}_{5} &=& \sigma_{0}\otimes (\hat{\bm{S}} \cdot \tilde{\bm{p}}) \\
    \mathcal{Q}_{7} &=& \sigma_{0}\otimes \hat{I}, \qquad & \mathcal{Q}_{8} &=& i\sigma_{2}\otimes (\hat{\bm{S}} \cdot \tilde{\bm{p}}).
  \end{array}
\end{equation}
These symmetries, however, still contain the duality transformation
$\mathcal{Q}_{2}$, which means that the medium is dual-symmetric and
supports the conservation of the electromagnetic helicity
\cite{Fernandez-Corbaton2013} and, as a consequence, optical zilches.

Definition of the optical chirality density in chiral media requires
some attention. This situation is similar to the definition of the
electromagnetic energy density \cite{Fedorov1976}. It can be
demonstrated that the chirality density in the medium with the
constituent relations (\ref{con1}) and (\ref{con2}) can be introduced
in the following way
\begin{equation}
  \rho_{\chi} = \frac{\varepsilon\varepsilon_{0}}{2} \bm{B}^{*}\cdot\frac{\partial \bm{E}}{\partial t}
  - \frac{\mu\mu_{0}}{2} \bm{D}^{*}\cdot\frac{\partial \bm{H}}{\partial t},
\end{equation}
which provides continuity of the chirality flow in spatially
inhomogeneous medium \cite{Proskurin2017}.

In order to understand the physical meaning of $\rho_{\chi}$, let us
look at energy absorption in a dissipative gyrotropic medium with
the constituent relations (\ref{con1}) and (\ref{con2}). As was
demonstrated in Ref.~\cite{Tang2010}, the electromagnetic energy
absorption rate in this case has an asymmetric part, which has
opposite signs for left and right polarized electromagnetic waves. This
part is proportional the product between the chirality of the
material, given by the imaginary part of $\varkappa$, and the
chirality density of the electromagnetic field $\rho_{\chi}$. The flow
of optical chirality in Eq.~(\ref{jchi}), in this situation, can be
associated with the asymmetric components of the electromagnetic
forces in the medium, which can be used, for example, for optical
separation of chiral molecules \cite{Canaguier-Durand2013}.

In the next section, we will show how these arguments can be
generalized to spin excitations in antiferromagnetic
materials. Similar to the results of this section, the symmetry
analysis will play a principal role in our discussion.

\section{Spin-wave chirality in antiferromagnetic insulators}
\label{sec:3}
The symmetry analysis developed in the previous section for Maxwell's
equations can be generalized to other dynamical systems.  Here, we
develop such generalization for spin-wave excitations in an
antiferromagnetic insulator.  A key observation that helps us to draw
the analogy between spin-wave dynamics and electrodynamics
is that the antiferromagnetic spin waves can be also characterized by
two polarization states.  This stems from the fact that the
magnetization dynamics in antiferromagnets involves two coupled
magnetic sublattices.  We, therefore, examine the
symmetry transformation in the extended space that includes three-dimensional rotations and transformations between the sublattices,
in order to find an algebra of nongeometric symmetries for
spin waves equivalent to that of the electrodynamics.

\subsection{Equations of motion for antiferromagnetic spin waves}
We start our discussion with a simple case of a collinear
antiferromagnet with two equivalent magnetic sublattices
$\bm{M}_{1}(t,\bm{r})$ and $\bm{M}_{2}(t,\bm{r})$.  The energy for
such antiferromagnet can be written in the following form
\begin{eqnarray}
  W = \int d^{3}r \left[
  \frac{\alpha}{2} \left(
  \frac{\partial \bm{M}_{1}}{\partial x_{\mu}} \cdot \frac{\partial \bm{M}_{1}}{\partial x_{\mu}} +
  \frac{\partial \bm{M}_{2}}{\partial x_{\mu}} \cdot \frac{\partial \bm{M}_{2}}{\partial x_{\mu}} \right)
  + \alpha' \frac{\partial \bm{M}_{1}}{\partial x_{\mu}} \cdot \frac{\partial \bm{M}_{2}}{\partial x_{\mu}}
  \right.  \nonumber
  \\ 
  \left.
  + \frac{\delta}{2} \bm{M}_{1} \cdot \bm{M}_{2} - \frac{\beta}{2} \left(
  \left(\bm{M}_{1} \cdot \bm{n}\right)^{2} + \left(\bm{M}_{2} \cdot \bm{n}\right)^{2} \right)
  \right], \label{eq:W}
\end{eqnarray}
where $\alpha$, $\alpha'$, and $\delta$ are the antiferromagnetic
exchange parameters and $\beta > 0$ describes the uniaxial magnetic
anisotropy with $\bm{n}$ being the unit vector along the anisotropy
axis \cite{Akhiezer1968}.  In the ground state, the anisotropy
stabilizes a uniform magnetic ordering along $\bm{n}$ where two
sublattices compensate each other, $\bm{M}_{1} = -\bm{M}_{2}$, so that
the total magnetization vanishes.

In the semi-classical limit, magnetization dynamics are described by
the Landau-Lifshitz-Gilbert equations of motion
\begin{equation}
  \label{eq:LLG}
  \frac{\partial \bm{M}_{i}}{\partial t} =
  \gamma \bm{M}_{i} \times \bm{H}_{i}^{\mathrm{eff}}
  - \eta \bm{M}_{i} \times \frac{\partial \bm{M}_{i}}{\partial t},
  \quad (i = 1,2),
\end{equation}
where $\gamma$ is the gyromagnetic ratio, $\bm{H}_{i}^{\mathrm{eff}} = -\delta W/\delta\bm{M}_{i}$ is the
effective field acting on the magnetization on the $i$th sublattice
and $\eta$ is the Gilbert damping that takes dissipation into account
\cite{Akhiezer1968}.

\begin{figure}[t]
	\sidecaption[t]
	\includegraphics[scale=.35]{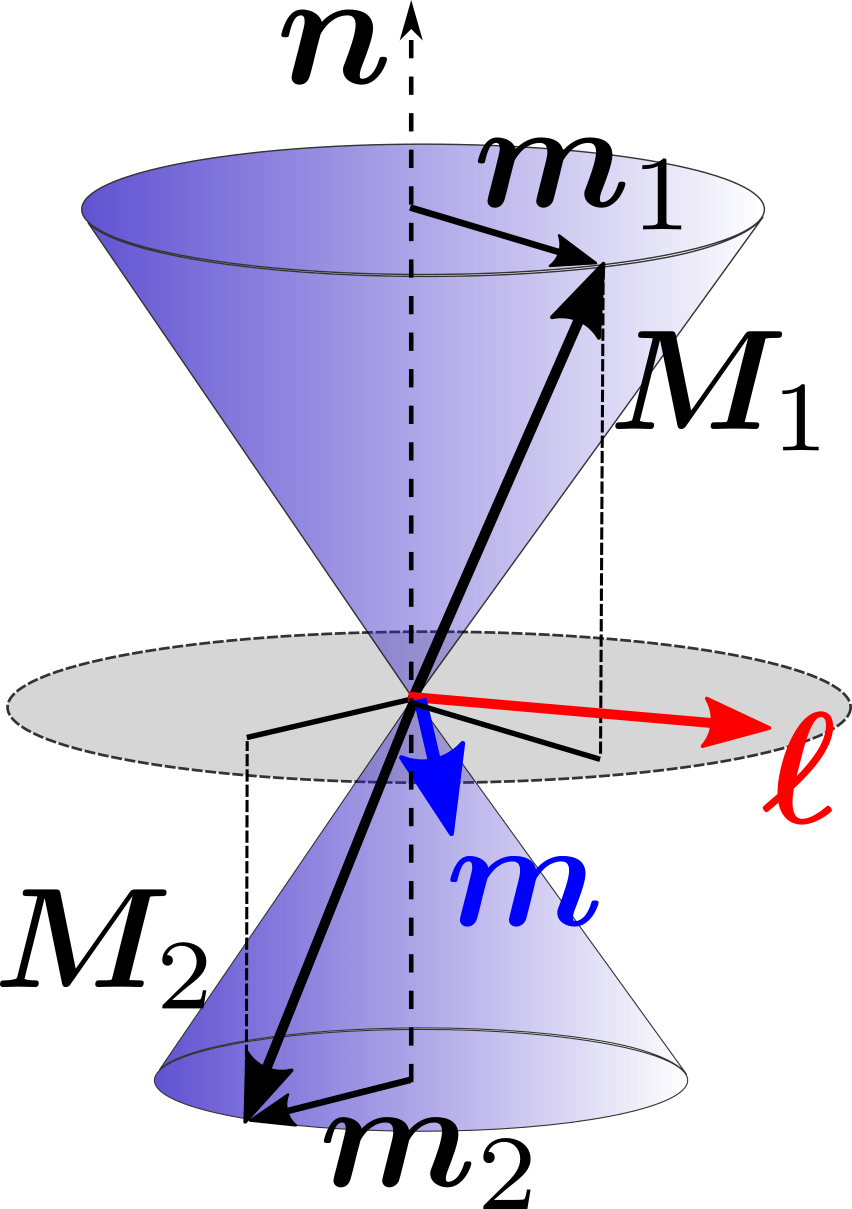}
	\caption{Sublattice magnetizations $\bm{M}_{1}$ and $\bm{M}_{2}$ precessing along the anisotropy axis $\bm{n}$; $\bm{m} = \bm{m}_{1} + \bm{m}_{2}$ is the resulting dynamic magnetization, and $\bm{l} = \bm{m}_{1} - \bm{m}_{2}$ shows the dynamic part of the antiferromagnetic vector.}
	\label{fig:1}
\end{figure}

For small excitations around the ground state configuration a linear
form of the Landau-Lifshitz-Gilbert equations of can be used.  This is
reached by breaking the sublattice magnetizations into static
$M_{s}\bm{n}$ and dynamic $\bm{m}_{i}(t, \bm{t})$ parts,
$\bm{M}_{i} = (-1)^{i+1}M_{s}\bm{n} + \bm{m}_{i}$, and keeping only
the linear terms in $\bm{m}_{i}$ in the resulting equations of motion ($M_{s}$ denotes the saturation magnetization).  For
convenience, we transform $\bm{m}_{i}(\bm{r})$ to momentum space, such that
$\bm{m}_{i}(t, \bm{r}) = \int d^{3}p \exp(i\bm{p} \cdot \bm{r})
\bm{m}_{i\bm{p}}(t)$, and introduce the dynamic vectors of the
magnetization,
$\bm{m}_{\bm{p}} = \bm{m}_{1\bm{p}} + \bm{m}_{2\bm{p}}$, and
antiferromagnetism,
$\bm{l}_{\bm{p}} = \bm{m}_{1\bm{p}} - \bm{m}_{2\bm{p}}$, see Fig.~\ref{fig:1}. The resulting
linear system of the equations of motions is given by
\begin{eqnarray}
  \label{eq:eom1}
  \frac{\partial \bm{m}_{\bm{p}}}{\partial t} &=& -\varepsilon^{(l)}_{\bm{p}} \bm{n} \times \bm{l}_{\bm{p}}
                                                  + \eta \bm{n} \times \frac{\partial \bm{l}_{\bm{p}}}{\partial t}, \\ \label{eq:eom2}
  \frac{\partial \bm{l}_{\bm{p}}}{\partial t} &=& -\varepsilon^{(m)}_{\bm{p}} \bm{n} \times \bm{m}_{\bm{p}}
                                                  + \eta \bm{n} \times \frac{\partial \bm{m}_{\bm{p}}}{\partial t},
\end{eqnarray}
where
$\varepsilon^{(m)}_{\bm{p}} = \gamma M_{s}(\delta + \beta + (\alpha +
\alpha')p^{2})$ and
$\varepsilon^{(l)}_{\bm{p}} = \gamma M_{s}(\beta + (\alpha -
\alpha')p^{2})$.

For the equations of motion (\ref{eq:eom1}) and (\ref{eq:eom2}), it is
possible to find a representation that is similar to the
Silberstein-Bateman form of the Maxwell's equations
\cite{Proskurin2017b}. For this purpose, we introduce a vector column
$\psi(t,\bm{p}) = (\bm{m}_{\bm{p}}, \bm{l}_{\bm{p}})^{T}$, which
allows us to rewrite the equations of motion for the spin waves in the
form Eq.~(\ref{SBM}), where the matrix in the right-hand side is now
given by
\begin{equation}
  \label{eq:Hm}
  \mathcal{H}_{\mathrm{m}} = \left(
    \begin{array}{cc}
      0 & -\varepsilon_{\bm{p}}^{(l)} (\hat{\bm{S}} \cdot \bm{n}) \\
      -\varepsilon_{\bm{p}}^{(m)} (\hat{\bm{S}} \cdot \bm{n}) & 0
    \end{array}
  \right).
\end{equation}
Here, we omit damping terms, which we discuss later. In this
form, the equations of motion for the spin waves resemble the
Maxwell's equations in a dispersive medium where the roles of the
electric permittivity and magnetic permeability is played by
$\varepsilon_{\bm{p}}^{(m)}$ and $\varepsilon_{\bm{p}}^{(l)}$.

The matrix in Eq.~(\ref{eq:Hm}) can be symmetrized by an appropriate
choice of the units that can be expressed in the form of the
transformation $\psi = \mathcal{N}\bar{\psi}$, where
$\mathcal{N} = \diag([\varepsilon_{\bm{p}}^{(m)}]^{-1/2},
[\varepsilon_{\bm{p}}^{(l)}]^{-1/2})$.  In the symmetric units, the
equation of motion for the antiferromagnetic spin waves is written as
\begin{equation}
  \label{eq:H0}
  i\frac{\partial \bar{\psi}(t,\bm{p})}{\partial t} =
  \mathcal{H}_{0}(\bm{p}) \bar{\psi}(t,\bm{p}),
\end{equation}
where the matrix on the right-hand side becomes symmetric
\begin{equation}
  \label{eq:H01}
  \mathcal{H}_{0}(\bm{p}) = \left(
    \begin{array}{cc}
      0 & -\omega_{\bm{p}}(\hat{\bm{S}} \cdot \bm{n}) \\
      -\omega_{\bm{p}}(\hat{\bm{S}} \cdot \bm{n}) & 0
    \end{array}
  \right) =  -\omega_{\bm{p}}
  \sigma_{1} \otimes (\hat{\bm{S}} \cdot \bm{n}),
\end{equation}
with
$\omega_{\bm{p}} =
\sqrt{\varepsilon_{\bm{p}}^{(m)}\varepsilon_{\bm{p}}^{(l)}}$.

This expression has a structure similar to $\mathcal{H}(\bm{p})$ in
Eq.~(\ref{eq:Hem}) for the Maxwell's equations.  The important
difference between $\mathcal{H}_{0}$ and $\mathcal{H}$ comes
from their properties under spatial inversion ($\mathcal{P}$) and time-reversal ($\mathcal{T}$) transformations. For example, in the case of
the time-reversal transformation, $ \phi(t, \bm{p}) $ in Eq.~(\ref{SBM})
transforms as
$\mathcal{T}\phi(t,\bm{p}) \to \sigma_{3} \phi(-t,\bm{p})$. The Pauli
matrix $\sigma_{3}$ appears on the right-hand side due to the
different transformation properties of the electric and magnetic field
with respect to $\mathcal{T}$.  In contrast, both components of
$\psi(t,\bm{p})$ are odd under $\mathcal{T}$, so that
$\mathcal{T}\psi(t,\bm{p}) \to -\psi(-t,\bm{p})$. This means that if we
want to transform from the spin wave dynamics to the electrodynamics,
we should replace $ \sigma_1 $ in Eq.~(\ref{eq:H01}) with
$ \sigma_{2} = i \sigma_{1} \sigma_{3} $ to ensure correct
properties under the $\mathcal{P}\mathcal{T}$ transformations.

\subsection{Nongeometric symmetries for spin-wave dynamics}
Formal analogy between the equations of motion for the
antiferromagnetic spin waves and the Maxwell's equations enables us to
generalize the concept of nongeometric symmetries. We may ask a
question about all the transformations
$\bar{\psi}(t, \bm{p}) \to \bar{\psi}'(t, \bm{p})$ that leave
the equation of motion (\ref{eq:H0}) invariant.

In order to find all such symmetries, one can repeat the steps of
Section~\ref{sec:2.1.1}.  First, we have to transform to the basis
where $\mathcal{H}_{0}(\bm{p})$ is diagonal. For this purpose, we make
a unitary transformation
$\bar{\psi} = \mathcal{U}_{\mathrm{m}}\tilde{\psi}$, where the
transformation matrix,
$\mathcal{U}_{\mathrm{m}} = U_{1} \otimes \hat{U}_{\Lambda}$, is given
by the rotation matrix to the helicity basis in Eq.~(\ref{UL}) (where
$\bm{p}$ is replaced by $\bm{n}$) combined with the $SU(2)$ rotation
in the subspace of $\bm{m}_{\bm{p}}$ and $\bm{l}_{\bm{p}}$
\begin{equation}
  \label{eq:U2m}
  U_{1} = \frac{1}{\sqrt{2}}  \left(
    \begin{array}{cc}
      1 & 1 \\
      -1 & 1
    \end{array}      \right).
\end{equation}

The resulting equation of motion for $\tilde{\psi}$ is given by
Eq.~(\ref{eq:H0}) with the diagonal matrix on the right-hand side
\begin{equation}
  \label{eq:tH0}
  \tilde{\mathcal{H}}_{0}
  = \mathcal{U}_{\mathrm{m}}^{\dag} \mathcal{H}_{0} \mathcal{U}_{\mathrm{m}}
  = \diag(-\omega_{\bm{p}}, \omega_{\bm{p}}, 0, \omega_{\bm{p}}, -\omega_{\bm{p}}, 0).
\end{equation}
This describes two antiferromagnetic spin waves with an energy
dispersion $\omega_{\bm{p}}$ degenerate with respect to the two
polarization states.  In an antiferromagnet, magnetization precession
is locked in the real space to the direction of $\bm{n}$, so that
these polarization states correspond to left and right circular
polarizations along the anisotropy axis.  This is in contrast
to electrodynamics, where we deal with real helicity --- precession around the direction of wave vector $\bm{p}$.

Secondly, we have to find all the matrices $\mathcal{Q}$ that commute
with $\tilde{\mathcal{H}}_{0}$, which can be done precisely in the
same way as in Eq.~(\ref{eq:Q1}).  It should be mentioned that in the
region $(\alpha - \alpha')p^{2} \gg \beta$, antiferromagnetic spin
waves have almost linear dispersion, $\omega_{\bm{p}} = c_{s}p$,
where the velocity is given by
$c_{s} = \gamma M_{s}\sqrt{\delta(\alpha - \alpha')}$.  This fact
gives them the appearance similar to the electromagnetic waves.
However, we emphasize that the linear dispersion is not essential for
our symmetry analysis.

What is important is that the eigenvalues of $\tilde{\mathcal{H}}_{0}$ are
degenerate. This fact allows us find the eight-dimensional algebra of
the symmetry transformations, which is isomorphic to invariance
algebra of the Maxwell's equations. The generators of this algebra can
be chosen as follows
\begin{equation}
  \label{eq:Qm}
  \begin{array}{rclrcl}
    \mathcal{Q}_{1} &=& i\sigma_{2} \otimes (\hat{\bm{S}} \cdot \bm{n})\hat{D}, \qquad &
                                                                                         \mathcal{Q}_{2} &=&  \sigma_{1} \otimes \hat{I},        \\
    \mathcal{Q}_{3} &=&  \sigma_{3} \otimes (\hat{\bm{S}} \cdot \bm{n})\hat{D}, \qquad &
                                                                                         \mathcal{Q}_{4} &=& i\sigma_{2} \otimes \hat{D},        \\
    \mathcal{Q}_{5} &=&  \sigma_{0} \otimes (\hat{\bm{S}} \cdot \bm{n}), \qquad &
                                                                                  \mathcal{Q}_{6} &=&  \sigma_{3} \otimes \hat{D},        \\
    \mathcal{Q}_{7} &=&  \sigma_{0} \otimes \hat{I}, \qquad &
                                                              \mathcal{Q}_{8} &=&  \sigma_{1} \otimes (\hat{\bm{S}} \cdot \bm{n}),        \\
  \end{array}
\end{equation}
where
$\hat{D} = 2[(\hat{\bm{S}} \cdot \bm{n}_{\perp})^{2} -
\hat{I}_{3}\bm{n}_{\perp}^{2}]/n_{\perp}^{2} - (\hat{\bm{S}} \cdot
\bm{n})^{2}$, $\hat{I}_{3} = \diag(0,0,1)$, and
$\bm{n}_{\perp} = (n_{1}, n_{2}, 0)$.  The interpretation of these
basis elements is similar to that in Eq.~(\ref{eq:Q}). We have the
unit element $\mathcal{Q}_{7}$, $\mathcal{Q}_{8}$ up to the factor of
$\omega_{\bm{p}}$ coincides with $\mathcal{H}_{0}(\bm{p})$ and,
therefore, commutes with all the other basis elements, and
$\mathcal{Q}_{5}$ generates rotations along $\bm{n}$.

Remarkably, $\mathcal{Q}_{2}$ plays a role of the duality
transformation of the electromagnetic field. It generates a continuous
symmetry transformation, the Bogolyubov's rotation, in the subspace of $\bm{m}_{\bm{p}}$
and $\bm{l}_{\bm{p}}$
\begin{eqnarray}
  \bm{m}_{\bm{p}} & \to & \bm{m}'_{\bm{p}} = \bm{m}_{\bm{p}} \cosh \theta + \sqrt{\frac{\varepsilon_{\bm{p}}^{(l)}}{\varepsilon_{\bm{p}}^{(m)}}} \bm{l}_{\bm{p}} \sinh \theta, \\
  \bm{l}_{\bm{p}} & \to & \bm{l}'_{\bm{p}} = \bm{l}_{\bm{p}} \cosh \theta + \sqrt{\frac{\varepsilon_{\bm{p}}^{(m)}}{\varepsilon_{\bm{p}}^{(l)}}} \bm{m}_{\bm{p}} \sinh \theta,
\end{eqnarray}
which leaves Eqs.~(\ref{eq:eom1}) and (\ref{eq:eom2}) invariant for
any real parameter $\theta$. Similar to the electrodynamics, we have
an algebraic property
$\mathcal{Q}_{2}\mathcal{Q}_{2} = \mathcal{Q}_{8}$, which establishes
a relation between the duality, the rotation symmetry along $\bm{n}$, and
$\partial_{t}$.

\subsection{Conserving chirality of spin waves}
The existence of the symmetry transformations makes possible a
formulation of the conservation laws that correspond to these
symmetries. Conserving quantities can be expressed in terms of
bilinear forms similar to Eq.~(\ref{eq:cons})
\begin{equation}
  \label{eq:cons1}
  C = \frac{1}{2} \int d^{3} p \psi^{\dag}(t, \bm{p}) \rho \mathcal{Q} \psi(t,\bm{p}),
\end{equation}
where $\mathcal{Q}$ is a symmetry transformation, which can be
expressed as a linear combination of $\mathcal{Q}_{i}$
($i = 1, \ldots, 8$), and the measure
$\rho = \diag(\varepsilon_{\bm{p}}^{(m)}, \varepsilon_{\bm{p}}^{(l)})$
is necessary for transforming from the symmetric representation of the
equations of motions in Eqs.~(\ref{eq:H0}) and (\ref{eq:H01}) to the
original units.

The conservation law for spin-wave chirality can be formulated
similar to the expression for the optical zilch in
Section~\ref{sec:2.1.3}.  Since the rotation symmetry is preserved
only along the direction of $\bm{n}$, we take the component of the
spin wave momentum along this direction
$\bm{p}_{n} = (\bm{p} \cdot \bm{n})\bm{n}$, and apply the conservation
law in Eq.~(\ref{eq:cons1}) for the symmetry transformation
$p_{n}\mathcal{Q}_{5} = (\hat{\bm{S}} \cdot \bm{p}_{n})$. As a result,
the expression for conserving spin-wave chirality is
given by
\begin{equation}
  \label{Cchim}
  C_{\chi}^{(\mathrm{m})} = \frac{i}{2} \int d^{3}p \left[
    \varepsilon_{\bm{p}}^{(m)} \bm{m}^{*}_{\bm{p}} \cdot (\bm{p}_{n} \times \bm{m}_{\bm{p}}) +
    \varepsilon_{\bm{p}}^{(l)} \bm{l}^{*}_{\bm{p}} \cdot (\bm{p}_{n} \times \bm{l}_{\bm{p}})
  \right],
\end{equation}
which is a direct analogue of the Lipkin's zilch for the
electromagnetic field. In real space, the chirality density for
spin waves can be written as
\begin{equation}
  \label{eq:rhom}
  \rho_{\chi}^{(\mathrm{m})}(t,\bm{r}) =
  \frac{1}{2}\left(  \nabla_{n} \bm{m} \cdot \frac{\partial \bm{l}}{\partial t} +
    \nabla_{n} \bm{l} \cdot \frac{\partial \bm{m}}{\partial t} \right),
\end{equation}
where $\nabla_{n} = \bm{\nabla} \cdot \bm{n}$.

Physical meaning of $C_{\chi}^{(\mathrm{m})}$ becomes clear if we
rewrite the expression (\ref{Cchim}) in terms of circularly polarized
magnon operators.  In this case, total spin wave chirality is
determined by the difference between the number of left
($N_{\bm{p}}^{(R)}$) and right ($N_{\bm{p}}^{(R)}$) polarized magnons
\cite{Proskurin2017b}
\begin{equation}
  C_{\chi}^{(\mathrm{m})} = 2 \sum_{\bm{p}} p_{n}\omega_{\bm{p}}
  \left(N_{\bm{p}}^{(L)} - N_{\bm{p}}^{(R)}\right).
\end{equation}
Similar expression exists for the Lipkin's zilch written in terms of
the polarized photon modes \cite{ Coles2012}. For a monochromatic spin wave, $C_{\chi}^{(\mathrm{m})}$ becomes proportional to the spin angular momentum component along $\bm{n}$, which in terms of magnon number operators is given by $S^{(n)} = \sum_{\bm{p}}(N_{\bm{p}}^{(L)} - N_{\bm{p}}^{(R)})$ \cite{ Coles2012}.

\subsection{Spin-wave chirality in dissipative media}
By now, we have established that spin waves in antiferromagnets
can be characterized by the chiral invariant
$C_{\chi}^{(\mathrm{m})}$, which is analogous to the Lipkin's zilch in
optics.  Similar to the optical case, we may ask a question: how can
we make this chirality of the spin waves visible?  To answer this
question, we should look at the symmetries. Since
$C_{\chi}^{(\mathrm{m})}$ is a pseudoscalar that is odd under
$\mathcal{P}$ and even under $\mathcal{T}$, we have to break the same
symmetries inside the antiferromagnet following the idea discussed in Section~\ref{sec:2.2} for the light-matter interactions in chiral metamaterials.

Since our model in Eq.~(\ref{eq:W}) is not chiral, we should provide
some symmetry breaking mechanism.  One interesting possibility of such
mechanism that is relevant for spintronic applications is based on
electron spin current \cite{Proskurin2017b}.  The flow of spin angular
momentum is odd under the spatial inversion and even under the time
reversal transformation, therefore, its interaction with
antiferromagnetic spin waves is able to provide the necessary symmetry
breaking.

\begin{figure}[t]
	\centerline{\includegraphics[scale=.325]{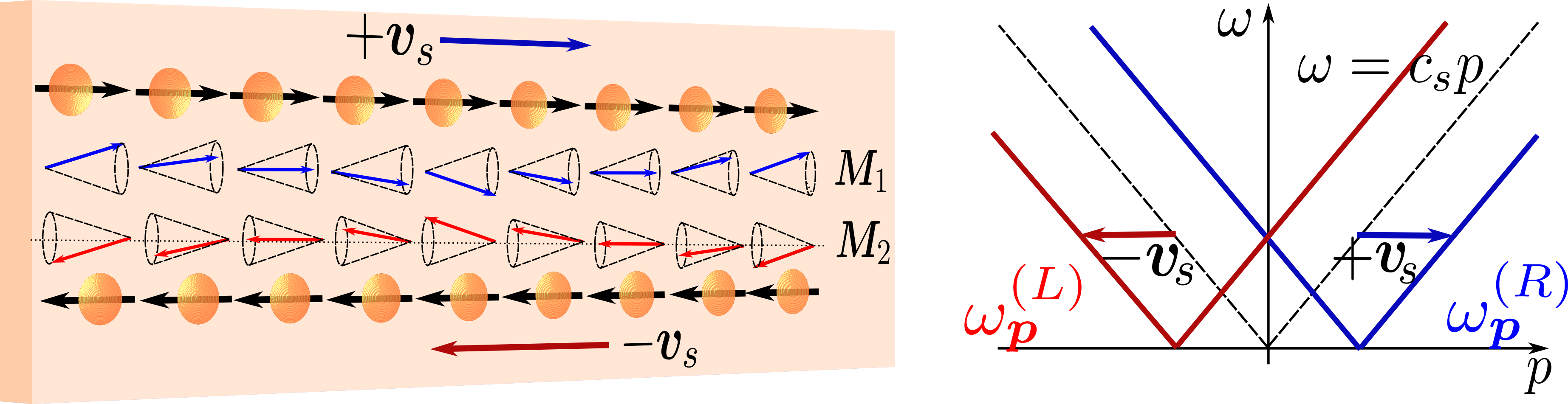}}
	\caption{Schematic picture of a pure spin current inside an antiferromagnet.  Spin majority (minority) electrons moving with the velocity $+v_{s}$ ($-v_{s}$) create adiabatic spin torque applied to $\bm{M}_{1}$ ($\bm{M}_{2}$).  These torques Doppler shift the energy dispersion of the left, $ \omega_{\bm{p}}^{(L)} $, and right, $ \omega_{\bm{p}}^{(R)} $, polarized modes in the opposite directions lifting the degeneracy between magnons of different polarizations.}
	\label{fig:2}
\end{figure}

The microscopic mechanism beyond this symmetry breaking is as follows.
Let us consider an electron spin current flowing along the magnetic
ordering direction $\bm{n}$, which can be injected into an
antiferromagnetic insulator film by a proximity effect or can be created in
bulk metallic antiferromagnets.  A pure spin current consists of a
number of spin majority electrons ($\uparrow$) polarized along
$\bm{n}$ flowing with the velocity $\bm{v}_{s}$ parallel to $\bm{n}$
balanced by the same amount of spin minority electrons ($\downarrow$)
moving with the velocity $-\bm{v}_{s}$, so that the net electric
charge transport is zero.  Since the spin-wave dynamics
is slow with respect to that of the electrons, the latter are able to
exert a spin transfer torque on the magnetization dynamics via the
Zhang-Li mechanism \cite{Zhang2004}. If the local $s$-$d$ interactions
between the electrons and sublattice magnetizations are in the
exchange dominant regime \cite{Yamane2016}, which means that we can
neglect the intersublattice electron scattering, the spin majority
(minority) electrons couple mostly to $\bm{M}_{1}$ ($\bm{M}_{2}$)
sublattice magnetization.  In this situation, the spin-$\uparrow$ electrons
create the spin transfer torque acting mostly on the magnetization
$\bm{M}_{1}$
\begin{equation}
  \bm{\mathfrak{T}}_{1} = -\frac{1}{M_{s}^{2}} \bm{M}_{1} \times
  (\bm{M}_{1} \times (\bm{v_{s} \cdot \bm{\nabla}}) \bm{M}_{1})
  -\frac{\xi}{M_{s}} \bm{M}_{1} \times (\bm{v}_{s} \cdot \bm{\nabla}) \bm{M}_{1},   
\end{equation}
where the first (second) term is the adiabatic (non-adiabatic) torque
component, and $\xi \lesssim 1$ is the dimensionless parameter
\cite{Zhang2004, Yamane2016}.  At the same time, spin-$\downarrow$
electron flow produce the spin transfer torque
$\bm{\mathfrak{T}}_{2} = -\bm{\mathfrak{T}}_{1}$ applied to
$\bm{M}_{2}$. Therefore, a pure spin current
in the exchange dominant regime of the electron-spin interaction is
able to create a pair of equal anti-parallel spin transfer torques
$\bm{\mathfrak{T}}_{1}$ and $\bm{\mathfrak{T}}_{2}$ acting on magnetizations $\bm{M}_{1}$ and $\bm{M}_{2}$ respectively, as schematically shown in Fig.~\ref{fig:2}.

\subsubsection{Doppler shift from a pure spin current}
The Landau-Lifshitz-Gilbert equations of motion for the magnetizations
in the presence of the spin-transfer torques are written as follows
\begin{eqnarray}
  \label{eom5}
  \frac{\partial \bm{M}_{1}}{\partial t} &=&
                                             \gamma \bm{M}_{1} \times \bm{H}_{1}^{\mathrm{eff}}
                                             + \eta \bm{M}_{1} \times \frac{\partial \bm{M}_{1}}{\partial t}
                                             - \frac{v_{s}}{M_{s}^{2}} \bm{M}_{1} \times (\bm{M}_{1} \times \nabla_{n}\bm{M}_{1}), \\
  \label{eom6}
  \frac{\partial \bm{M}_{2}}{\partial t} &=& \gamma \bm{M}_{2} \times \bm{H}_{2}^{\mathrm{eff}} + \eta \bm{M}_{2} \times \frac{\partial \bm{M}_{2}}{\partial t} + \frac{v_{s}}{M_{s}^{2}} \bm{M}_{2} \times (\bm{M}_{2} \times \nabla_{n}\bm{M}_{2}), 
\end{eqnarray}
where we neglect non-adiabatic contribution to the spin
torque. Taking into account that $|\bm{M}_{i}| = M_{s}$ ($i = 1,2$),
these expressions can be rewritten as follows
\begin{equation}
  \left(\frac{\partial}{\partial t} \mp v_{s}\nabla_{n}\right) \bm{M}_{i} = \gamma \bm{M}_{i} \times \bm{H}_{i}^{\mathrm{eff}} + \eta \bm{M}_{i} \times \frac{\partial \bm{M}_{i}}{\partial t},
\end{equation}
where the upper (lower) sign is for $i=1$ ($i=2$). This expression
shows that the role of the adiabatic spin transfer torque is to
produce a Doppler shift of the spin waves by the velocity
$v_{s}$. This effect is well-known for ferromagnetic and
antiferromagnetic spin waves when the Doppler shift is caused by a
spin polarized electric current \cite{Yamane2016, Vlaminck2008,
  Swaving2011}.  In our case, the pure spin current produces two
Doppler shifts in the opposite directions for the magnetization
dynamics on each sublattice.

By solving the equations of motion (\ref{eom5}) and (\ref{eom6}), it
is possible to show that in the presence of the spin current, the
degeneracy between left and right polarizations in the dispersion
relations for the spin waves propagating along $\bm{n}$ becomes
lifted, and it can be approximated as follows
\cite{Proskurin2017b}
\begin{eqnarray}
  \omega_{\bm{p}}^{(R)} &=& c_{s}|p - p_{s}| + i\eta (\Delta_{s} - pv_{s}), \\
  \omega_{\bm{p}}^{(L)} &=& c_{s}|p + p_{s}| + i\eta (\Delta_{s} + pv_{s}),
\end{eqnarray}
where $p_{s} = \gamma M_{s}v_{s}\delta/(2c_{s}^{2})$,
$\Delta = \gamma M_{s}\delta /2$, and $p \gg p_{s}$ is the wave vector
of the spin waves along $\bm{n}$, see Fig.~\ref{fig:2}.

This effect is in contrast to the
Doppler shift from a spin polarized current where both modes are
shifted in the same direction so that the degeneracy holds \cite{Yamane2016}. The imaginary parts
of the frequencies $\omega_{\bm{p}}^{(R)}$ and $\omega_{\bm{p}}^{(L)}$
also have contributions from the spin current of the opposite
signs for the waves with left and right polarizations. This can be
considered as a spin-current-induced circular dichroims of spin
waves, which occurs at the characteristic length scale
$\ell_{\mathrm{CD}} = c_{s}/(\eta v_{s}p )$.

Interestingly, the effect of spin current on the spin waves in the
linear approximation is analogous to the existence of the additional
Dzyaloshinskii-Moriya interaction (DMI) term in the antiferromagnetic energy in
Eq.~(\ref{eq:W})
\begin{equation}
  \label{eq:Wdmi}
  W_{\mathrm{DMI}} = \frac{v_{s}}{2\gamma M_{s}} \int d^{3} r \left[ \bm{m}_{1} \cdot (\bm{\nabla}_{n} \times \bm{m}_{1}) + \bm{m}_{2} \cdot (\bm{\nabla}_{n} \times \bm{m}_{2})  \right],
\end{equation}
between the magnetizations on the same sublattices.

\subsubsection{Asymmetric energy absorption}
Let us now look at the spin-wave energy absorption.  The 
dissipation rate for the magnetization dynamics can be expressed
through the Rayleigh dissipation function
\begin{equation}
  \label{eq:dWdt}
  \frac{d W}{d t} = -\frac{\eta}{\gamma} \int d^{3}r \left[
    \left(\frac{\partial \bm{M}_{1}}{\partial t}\right)^{2}
    +\left(\frac{\partial \bm{M}_{2}}{\partial t}\right)^{2} \right].
\end{equation}
According to the equations of motion (\ref{eom5}) and (\ref{eom6}), in
the presence of the spin current we replace $\partial_{t}$ with
$\partial_{t} - v_{s}\nabla_{n}$ for $\bm{M}_{1}$ and with
$\partial_{t} + v_{s}\nabla_{n}$ for $\bm{M}_{2}$. The energy
dissipation rate in Eq.~(\ref{eq:dWdt}) in this case acquires the
asymmetric contribution proportional to $v_{s}$ that is written as
\begin{equation}
  \left(\frac{d W}{d t}\right)_{\chi} = \frac{2\eta v_{s}}{\gamma} \int d^{3}r
  \left(\nabla_{n} \bm{m}_{1} \cdot \frac{\partial \bm{m}_{1}}{\partial t} -
    \nabla_{n} \bm{m}_{2} \cdot \frac{\partial \bm{m}_{2}}{\partial t}
  \right).
\end{equation}
The expression in parentheses is nothing but the spin-wave chirality
density $\rho_{\chi}^{(\mathrm{m})}$ written in
terms of $\bm{m}_{1}$ and $\bm{m}_{2}$.

As a result, when a pure spin current is injected into an
antiferromagnet, the asymmetry in the spin-wave energy absorption rate
becomes proportional to the spin-wave chirality,
$(dW/dt)_{\chi} = 2\eta v_{s}\gamma^{-1} C_{\chi}^{\mathrm{m}}$.  This
result is a direct analogy with the result of Tang and Cohen
\cite{Tang2010} for the electromagnetic energy absorption rate in
chiral metamaterials, see Section~\ref{sec:2.2}.  In antiferromagnetic
materials, the microscopic mechanism beyond this phenomenon can be
based on the adiabatic spin transfer torque from a pure spin current,
or on the DMI between the same sublattices, which
breaks the inversion symmetry and lifts the degeneracy between the
left and right polarized magnon modes. In contrast to optical
metamaterials, where the asymmetry in light-matter interactions is related to structural chirality, the symmetry breaking
mechanism, which is based on the spin current, induces
chirality of the material in controllable way.  For a spin current
density $j_{s} \approx 10^{11}$~A/m$^{2}$ (in the electric units), we
obtain $v_{s} = \mu_{B}j_{s}/(eM_{s}) \approx 30$~m/s for
$M_{s} \approx 3.5 \times 10^{5}$~A/m. This parameter should be
compared to the typical velocity of the spin waves in
antiferromagnetic insulators $c_{s} \approx 10^{-4}$~m/s, which gives
$v_{s}/c_{s} \approx 10^{-3}$. The characteristic length of the magnon
circular dichroism, in this situation,
$\ell_{\mathrm{CD}} \approx 5$~mm for the magnon frequencies about
$1$~THz and $\eta \approx 10^{4}$. Curiously, the effective strength
of the DMI,
$D_{\mathrm{eff}} = \hbar v_{s} / (k_{B}a_{0})$ is about $0.5$~K ($a_{0}$ is
the lattice spacing), which is comparable to a typical DMI strength in
magnetic materials.

\section{Excitation of magnon spin photocurrents with polarized fields}
\label{sec:4}
Among the major goals of spintronics are generation of spin currents, their transmission over large distances, and conversion from one form to another because the spin angular momentum can be carried by different types of carriers.  Since magnons are able to carry spin angular momentum, spin excitations in low damping magnetic insulators are good candidates for being spin current mediators.  The absence of the net magnetization and the existence of two polarization states per magnon make antiferromagnetic insulators particularly suitable for applications as spin current conductors.  It was demonstrated that an introduction of a thin layer of the antiferromagnetic insulator can enhance the spin current transmission in interface systems \cite{Wang2014, Khymyn2016}.

Magnon spin currents in antiferromagnetic insulators can be excited by several methods.  For example, it can be done by pumping a magnon spin current from a neighboring ferromagnetic layer \cite{Wang2014}.  Thermal excitation of spin currents via the spin versions of the Seebeck and Nernst effects also has attracted considerable attention \cite{Seki2015, Rezende2016, Rezende2016a, Wu2016, Holanda2017}.  The latter is especially interesting in low-dimensional materials, where it is provided by topological terms in magnon dynamics \cite{Cheng2016a,Zyuzin2016,Shiomi2017}.

Optical control of spin states in antiferromagnetic insulators \cite{Satoh2010, Tzschaschel2017} is a feature in the emerging field of antiferromagnetic optospintronics \cite{Nemec2018}.  In this respect, it is an intriguing problem to investigate whether it is possible to find some sort of magnon photo-effect \cite{Proskurin2018a}.  Symmetry considerations suggest that this is indeed possible.  As we have already mentioned, spin currents satisfy the definition of true chirality \cite{Barron2004}, which can be directly seen from the conservation law for the $\mu$th component of the spin density
\begin{equation}
  \frac{\partial s^{\mu}(t,\bm{r})}{\partial t} + \bm{\nabla} \cdot \bm{j}^{\mu}(t,\bm{r}) = 0.
\end{equation}
Since $s^{\mu}(t,\bm{r})$ is $\mathcal{T}$ odd and $\mathcal{P}$ even, the spin current density $\bm{j}^{\mu}(t,\bm{r})$ has opposite transformation properties. As we have seen in Section~\ref{sec:2}, the electromagnetic field can be characterized by optical chirality $\rho_{\chi}(t,\bm{r})$ with the same transformations properties as $ \bm{j}^{\mu}(t,\bm{r})$.  Therefore, we may expect that by exposing an antiferromagnetic insulator to a circularly polarized electromagnetic field, we can excite a spin photocurrent, which direction should be determined by the helicity of light.

In this section, we will consider these arguments in detail, and show that this photo-excitation process requires the frequency of the electromagnetic field to be in the region of the antiferromagnetic resonance. We begin with a semiclassical theory. Nonlinear response and geometric effects in low dimensional materials are discussed at the end of this section.  First we consider an interesting phenomenon analogous to the \emph{Zitterbewegung} effect for magnons.

\subsection{Magnon spin currents in antiferromagnets}
\label{sec:4.1}
Equations (\ref{eq:eom1}) and (\ref{eq:eom2}) preserve rotation symmetry along the magnetic ordering direction that warrants conservation of the total angular momentum component along $\bm{n}$.  From these equations, the time evolution of the $n$th component of the magnetization $M^{(n)} = \frac{1}{2M_{s}}(m_{2}^{2} - m_{1}^{2})$ is written in the following form
\begin{eqnarray}
  \frac{\partial M^{(n)}(t,\bm{r})}{\partial t} =
  \frac{1}{4M_{s}}\sum_{\bm{p}\bm{q}}e^{-i\bm{q}\cdot\bm{r}}
  \bm{n} \cdot \left\lbrace \left(\varepsilon_{\bm{p}-\bm{q}}^{(l)} - \varepsilon_{-\bm{p}}^{(l)}\right)
  \left[ \bm{l}^{*}_{\bm{p} - \bm{q}} \times \bm{l}_{\bm{p}} \right]
  \right.    \nonumber
  \\ 
  \left.
    +\left(\varepsilon_{-\bm{p}+\bm{q}}^{(m)} - \varepsilon_{\bm{p}}^{(m)}\right)
  \left[ \bm{m}^{*}_{\bm{p} - \bm{q}} \times \bm{m}_{\bm{p}} \right]
  \right\rbrace.
\end{eqnarray}
In the limit $\bm{q} \to 0$, this equation can be rewritten in the form of a continuity equation $\partial_{t}M_{\bm{q}}^{(n)} + i\bm{q} \cdot \bm{J}_{s}^{(n)} = 0$, where
\begin{equation}
  \label{eq:Js}
  \bm{J}_{s}^{(n)} = \frac{i}{4M_{s}} \sum_{\bm{p}} \left(\frac{\partial \varepsilon_{\bm{p}}^{(m)}}{\partial \bm{p}}\bm{m}^{*}_{\bm{p}} \cdot (\bm{n} \times \bm{m}_{\bm{p}})
    + \frac{\partial \varepsilon_{\bm{p}}^{(l)}}{\partial \bm{p}}\bm{l}^{*}_{\bm{p}} \cdot (\bm{n} \times \bm{l}_{\bm{p}})\right)
\end{equation}
is the total magnon spin current. This expression looks similar to our definition of the spin-wave chirality in Eq.~(\ref{Cchim}), especially if we consider the spin current flow along $\bm{n}$. However, as we shall see below, in contrast to magnon chirality, $\bm{J}_{s}^{(n)}$ does not obey any conservation law. It should be mentioned that the same expression for the spin current can be obtained directly from the antiferromagnetic Lagrangian using Noether's theorem (see Appendix).

It is interesting to discuss the analogy between antiferromagnetic magnon spin currents and charge currents in pseudo-relativistic Dirac materials.  In the latter case, it was demonstrated that interband effects make a significant contribution near the Dirac point and can explain, for example, the universal conductivity of graphene \cite{Katsnelson2006}. In the relativistic language, interband effects in the dynamics of an electron wave packet correspond to the \emph{Zitterbewegung}, or the trembling motion of an ultra-relativistic particle \cite{Katsnelson2006}. The \emph{Zitterbewegung} effect has also been proposed for antiferromagnetic magnons \cite{Wang2017}.  It can be easily understood by looking at the time evolution of $\bar{\psi}_{\bm{p}}(t)$ calculated from Eqs.~(\ref{eq:H0}) and (\ref{eq:H01})
\begin{equation}
  \label{eq:psi}
  \bar{\psi}_{\bm{p}}(t) = \frac{1}{2} \left\lbrace 
    \left[1 + \sigma_{1} \otimes (\hat{\bm{S}} \cdot \bm{n}) \right]e^{i\omega_{\bm{p}}t} +
    \left[1 - \sigma_{1} \otimes (\hat{\bm{S}} \cdot \bm{n}) \right]e^{-i\omega_{\bm{p}}t}
  \right\rbrace \bar{\psi}_{\bm{p}}(0),
\end{equation}
which is similar to the analogous equation for relativistic particles \cite{Katsnelson2006}.  This expression contains the off-diagonal elements responsible for the mixing of $\bm{m}_{\bm{p}}$ and $\bm{l}_{\bm{p}}$ components of $\bar{\psi}_{\bm{p}}$ while evolving in time.

By applying Eq.~(\ref{eq:psi}) to the time evolution of the spin current in Eq.~(\ref{eq:Js}),  we find that  the spin current has two contributions, $\bm{J}_{s}^{(n)}(t) = \bm{J}_{s0}^{(n)} + \bm{J}_{s1}^{(n)}(t)$. The first contribution is conserved part of the spin current. It does not depend on time and is proportional to the group velocity of magnons $v_{\bm{p}} = \partial \omega_{\bm{p}} / \partial \bm{p}$. In our matrix notations, it can be written as 
\begin{equation}
  \bm{J}_{s}^{(n)} = \frac{1}{4M_{s}} \sum_{\bm{p}} v_{\bm{p}} \bar{\psi}^{\dag}(0) (\hat{\bm{S}} \cdot \bm{n}) \bar{\psi}_{\bm{p}}(0).
\end{equation}
The second term in the spin current oscillates at the double frequency, and can be attributed to the  \emph{Zitterbewegung} of magnons
\begin{equation}
  \bm{J}_{s1}^{(n)}(t) = \frac{1}{16M_{s}} \sum_{\bm{p}} e^{2i\omega_{\bm{p}}t}
  \bm{K}_{\bm{p}} \bar{\psi}_{\bm{p}}^{\dag}(0) 
\left(
  \begin{array}{cc}
   (\hat{\bm{S}} \cdot \bm{n}) &  1 \\
 -1 & -(\hat{\bm{S}} \cdot \bm{n})
  \end{array}
\right)
\bar{\psi}_{\bm{p}}(0) + \mathrm{H.c.},
 \end{equation}
where 
\begin{equation}
\bm{K}_{\bm{p}} = \frac{1}{\omega_{\bm{p}}} \left( \varepsilon_{\bm{p}}^{(l)} \frac{\partial \varepsilon_{\bm{p}}^{(m)}}{\partial \bm{p}} -  \varepsilon_{\bm{p}}^{(m)} \frac{\partial \varepsilon_{\bm{p}}^{(l)}}{\partial \bm{p}}\right).
\end{equation}
The physical meaning of these terms becomes clear if we transform to the helicity basis, $ \tilde{\psi}_{\bm{p}} = (\tilde{\psi}^{(R)}_{\bm{p}}, \tilde{\psi}^{(L)}_{\bm{p}})^{T} $, where we have well-defined left and right polarized magnon modes, see Eqs.~(\ref{UL}), (\ref{eq:U2m}) and (\ref{eq:tH0}). In this basis, the first term is determined by the difference in numbers of magnons with opposite polarizations
\begin{equation}
  \label{eq:J00}
  \bm{J}_{s}^{(n)} = \frac{1}{4M_{s}} \sum_{\bm{p}} v_{\bm{p}} \left(\tilde{\psi}^{*(R)}_{\bm{p}}\tilde{\psi}^{(R)}_{\bm{p}} -  \tilde{\psi}^{*(L)}_{\bm{p}}\tilde{\psi}^{(L)}_{\bm{p}}\right),
\end{equation}
while the second term is purely off-diagonal and corresponds to the interband processes
\begin{equation}
  \label{eq:J10}
  \bm{J}_{s1}^{(n)}(t) = -\frac{1}{8M_{s}} \sum_{\bm{p}} \tilde{\psi}^{\dag}_{\bm{p}}(0)
  \left(
    \begin{array}{cc}
      0 & \bm{K}_{\bm{p}} \hat{S}^{z} e^{-2i\omega_{\bm{p}}\hat{S}^{z}t} \\
      \bm{K}_{\bm{p}} \hat{S}^{z} e^{2i\omega_{\bm{p}}\hat{S}^{z}t} & 0
    \end{array}
    \right)\tilde{\psi}_{\bm{p}}(0).
\end{equation}

It should be mentioned that the contribution of the oscillating term in total spin current may seem insignificant.  Indeed, in the theory the  spin Seebeck effect only the term given by Eq.~(\ref{eq:J00}) was taken into account in the definition of the spin current \cite{Rezende2016,Rezende2016a}.  In this case, the second term, which mixes magnons of different helicities, has vanishing contribution.  However, as we discuss below, such processes as the photo-excitation require both terms being considered with equal attention.  Moreover, the contribution of the second term in Eq.~(\ref{eq:J10}) may become dominant in low-dimensional systems where it may contain geometric phase effects.

\subsection{Photo-excitation of magnon spin currents}
\label{sec:4.2}
Let us now turn to a semi-classical theory of photo-excitation of magnon spin currents.  For this purpose, we add a magneto-dipole interaction between the magnetic field component of the electromagnetic wave $\bm{h}(t,\bm{r})$ and the magnetization of the antiferromagnet, so that the total energy is written as
\begin{equation}
  W_{t} = W - \int d^{3}r  (\bm{M}_{1} + \bm{M}_{2}) \cdot \bm{h}(t, \bm{r}),
\end{equation}
where $W$ is determined by Eq.~(\ref{eq:W}). In this case, Eq.~(\ref{eq:eom2}) acquires the additional term $-2\gamma M_{s} [\bm{n} \times \bm{h}_{\bm{p}}(t)]$, where $\bm{h}_{\bm{p}}(t)$ is the Fourier component of the magnetic field.  The system of equations of motion (\ref{eq:eom1}) and (\ref{eq:eom2}) can be easily solved by transforming the $\omega$-domain, which gives
\begin{eqnarray}
  \bm{m}_{\bm{p}}(\omega) & = & 2\gamma M_{s} \frac{\varepsilon_{\bm{p}}^{(l)} \bm{h}_{\bm{p}}(\omega)}{\omega_{\bm{p}}^{2} - \omega^{2}}, \\
  \bm{l}_{\bm{p}}(\omega) & = & 2i\gamma M_{s} \frac{\omega [\bm{n} \times \bm{h}_{\bm{p}}(\omega)]}{\omega_{\bm{p}}^{2} - \omega^{2}}.
\end{eqnarray}
The Gilbert damping can be phenomenologically introduced in these equations by considering complex parameters $\varepsilon_{\bm{p}}^{(\alpha)} \to \varepsilon_{\bm{p}}^{(\alpha)} - i\eta \omega$ ($\alpha = m,l$). Using the definition of the spin current in Eq.~(\ref{eq:Js}), we find the current excited by the magnetic field vector
\begin{equation}
  \label{eq:Js1}
  \bm{J}_{s}^{(n)} = i\gamma^{2}M_{s}\sum_{\bm{p}\omega}
  \frac{\varepsilon_{\bm{p}}^{(l)2} \partial_{\bm{p}}\varepsilon_{\bm{p}}^{(m)} + \omega^{2} \partial_{\bm{p}}\varepsilon_{\bm{p}}^{(l)}}{\left(\omega^{2} - \omega_{\bm{p}}^{2}\right)^{2}}
  \bm{h}_{\bm{p}}^{*}(\omega) \cdot [\bm{n} \times \bm{h}_{\bm{p}}(\omega)].
\end{equation}

\begin{figure}[t]
	\sidecaption[t]
	\includegraphics[scale=0.275]{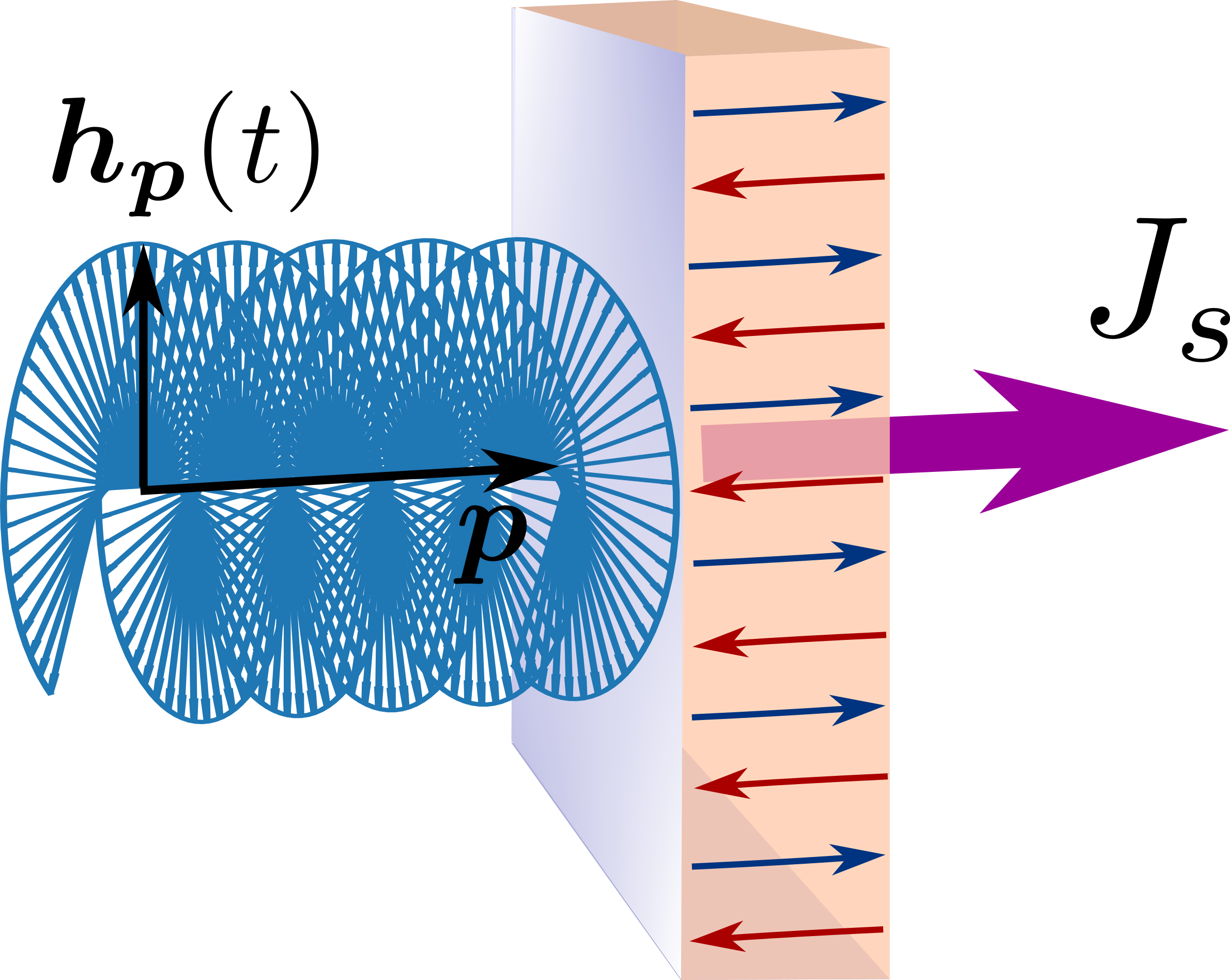}
	\caption{Schematic picture of the magnon photocurrent $\bm{J}_{s}^{(n)}$
          induced inside an antiferromagnet by the circularly polarized
          electromagnetic wave propagating along the direction of
          magnetic ordering.}
	\label{fig:3}
\end{figure}

This expression shows that the direct spin current excited by the electromagnetic wave is the second order effect in $\bm{h}_{\bm{p}}(\omega)$, and is determined by the asymmetric combination $\bm{h}^{*}_{\bm{p}} \times \bm{h}_{\bm{p}}$, so that the direction of the current is determined by helicity of the electromagnetic wave.  The effect is resonantly amplified near the antiferromagnetic resonance $\omega \approx \omega_{\bm{p}}$.

Photo-excitation of magnon spin currents in antiferromagnetic insulators shows some similarity with the circular photogalvanic effect in noncentrosymmetric metals \cite{Belinicher1980}.  In the latter case, a direct electric photocurrent is generated by the helical combination the electric-field vector of the electromagnetic wave, $\bm{E}^{*}(\omega) \times \bm{E}(\omega)$, so that the direction of the current is reversed whenever circular polarization of light is switched to the opposite.

In order to have further insight into magnon spin photocurrents, let us consider a quantum variant of our theory.

\subsection{Microscopic theory of magnon spin photocurrents}
The spin Hamiltonian for an antiferromagnetic insulator with two magnetic sublattices $A$ and $B$ can be written in the following form
\begin{equation}
  \label{eq:Hsp}
  \hat{H} = \sum_{ij} \frac{1}{2} \left(J_{ij} S_{i}^{(+)}S_{j}^{(-)} + J_{ij}^{*} S_{i}^{(-)}S_{j}^{(+)}\right) + 
  \sum_{ij} J'_{ij} S_{i}^{z}S_{j}^{z} - K \sum_{i} (S_{i}^{z})^{2},
\end{equation}
where  $J_{ij}$ and $J'_{ij}$ are the exchange interaction constants such as $\Re J_{ij} > 0$ and $J'_{ij} >0$ for the nearest neighboring sites on $A$ and $B$ sublattices, and $K \sim \beta a_{0}^{-3}$ is the magnetic anisotropy that stabilizes the antiferromagnetic ordering along the $z$ direction.  We do not specify any lattice configuration at this stage.  However, we note that $J_{ij}$ may have a complex phase factors in the presence of DMI.

The spin-wave approximation for the Hamiltonian (\ref{eq:Hsp}) is conveniently expressed by the Holstein–Primakoff transformation of the spin operators
\begin{equation}
  \begin{array}{rclrcl}
    S_{iA}^{(+)} &=& \sqrt{2S} a_{i}, \qquad & S_{iB}^{(+)} &=& \sqrt{2S} b_{i}^{\dag}, \\
    S_{iA}^{(-)} &=& \sqrt{2S} a_{i}^{\dag}, \qquad & S_{iB}^{(-)} &=& \sqrt{2S} b_{i}, \\
    S_{iA}^{z} &=& S - a_{i}^{\dag}a_{i}, \qquad & S_{iB}^{z} &=& -S + b^{\dag}b_{i}^{\dag}, \\
  \end{array}
\end{equation}
where $a_{i}$ and $b_{i}$ are boson operators at the $A$ and $B$ sublattice respectively, which satisfy boson commutation relations.  By transforming these operators to the reciprocal space, $a_{i} = \sum_{\bm{k}} \exp(i\bm{k}\cdot \bm{r}_{i}) a_{\bm{k}}$ and $b_{i} = \sum_{\bm{k}} \exp(i\bm{k}\cdot \bm{r}_{i}) b_{\bm{k}}$, we can rewrite Eq.~(\ref{eq:Hsp}) in the following form
\begin{equation}
\label{eq:Hmag}
\hat{H} = \sum_{\bm{k}} \left[A_{\bm{k}} \left(a_{\bm{k}}^{\dag}a_{\bm{k}} + b_{-\bm{k}}^{\dag}b_{-\bm{k}} \right)
+ B_{\bm{k}} a_{\bm{k}} b_{-\bm{k}} + B^{*}_{\bm{k}} a^{\dag}_{\bm{k}} b^{\dag}_{-\bm{k}}\right],
\end{equation}
where parameters $A_{\bm{k}}$ and $B_{\bm{k}}$ include microscopic details.  For example,  in the case when the exchange interactions are limited by the nearest neighboring sites so that $J_{ij} = J^{'}_{ij} = J_{1}$, we obtain $A_{\bm{k}} = 2KS + ZJ_{1}S$ and $B_{\bm{k}} = J_{1}S\sum_{\bm{\delta}} \exp(-i\bm{k} \cdot \bm{\delta})$, where $\bm{\delta}$ connects a site on the $A$ sublattice with its $Z$  nearest neighboring sites on the $B$ sublattice.

\subsubsection{Magnon spin currents: quantum version}
The expression for a magnon spin current can be derived following the same steps as in Sec.~\ref{sec:4.1}.  Considering the equation of motion for the $z$ component of the local spin density, $n(\bm{r}_{i}) = b^{\dag}_{i}b_{i} - a^{\dag}_{i}a_{i}$, we find the total magnon spin current
\begin{equation}
\hat{\bm{J}}_{s} = \sum_{\bm{k}} \left[\frac{\partial A_{\bm{k}}}{\partial \bm{k}} \left(a_{\bm{k}}^{\dag}a_{\bm{k}} + b_{-\bm{k}}^{\dag}b_{-\bm{k}} \right)
+ \frac{\partial B_{\bm{k}}}{\partial \bm{k}} a_{\bm{k}} b_{-\bm{k}} + \frac{\partial B^{*}_{\bm{k}}}{\partial \bm{k}} a^{\dag}_{\bm{k}} b^{\dag}_{-\bm{k}}\right].
\end{equation}
This expression can be conveniently written in the matrix form 
\begin{equation}
\hat{\bm{J}}_{s} = \sum_{\bm{k}} \chi_{\bm{k}}^{\dag} \frac{\partial \mathcal{H}_{\bm{k}}}{\partial \bm{k}} \chi_{\bm{k}},
\end{equation}
where we introduced $\chi_{\bm{k}} = \left(\begin{array}{c} a_{\bm{k}} \\ b^{\dag}_{-\bm{k}} \end{array}\right)$ and
$
\mathcal{H}_{\bm{k}} = \left(
\begin{array}{cc}
A_{\bm{k}} & B^{*}_{\bm{k}} \\
B_{\bm{k}} & A_{\bm{k}}
\end{array}
\right)
$.
Note that in this representation, $\chi_{\bm{k}}$ does not satisfy boson communication relations; instead one has $[\chi_{\bm{k}}, \chi_{\bm{k}'}^{\dag}] = \sigma_{z} \delta_{\bm{k},\bm{k}'}$, which should be kept in mind.

Let us find how $\hat{\bm{J}}_{s}$ transforms under the Bogolyubov's transformation that preserves boson commutation relations of magnon operators.  In the matrix form, this transformation is expressed as $\chi_{\bm{k}} = U_{\bm{k}}\tilde{\chi}_{\bm{k}}$, where the transformation matrix is determined by two real parameters $\theta_{\bm{k}}$ and $\phi_{\bm{k}}$:
\begin{equation}
U_{\bm{k}} = 
\left(
\begin{array}{cc}
\cosh \theta_{\bm{k}} e^{i\phi_{\bm{k}}} & -\sinh \theta_{\bm{k}} \\
-\sinh \theta_{\bm{k}} & \cosh \theta_{\bm{k}} e^{-i\phi_{\bm{k}}}
\end{array}
\right).
\end{equation}
Since the definition of spin current involves $\partial_{\bm{k}}$, its transformation properties invoke covariant derivatives with respect to $U_{\bm{k}}$.  Explicit calculations show that in an arbitrary basis 
\begin{equation}
\label{eq:Jsmt}
\hat{\bm{J}}_{s} = \sum_{\bm{k}} \tilde{\chi}_{\bm{k}}^{\dag} \frac{\partial \tilde{\mathcal{H}}_{\bm{k}}}{\partial \bm{k}} \tilde{\chi}_{\bm{k}} - \frac{\partial \hat{\bm{A}}}{\partial t},
\end{equation}
where $\tilde{\mathcal{H}}_{\bm{k}} = U^{\dag}_{\bm{k}} \mathcal{H}_{\bm{k}} U_{\bm{k}}$ is the Hamiltonian in the transformed basis, and $\hat{\bm{A}} = \sum_{\bm{k}} \tilde{\chi}_{\bm{k}}^{\dag} \mathcal{A}_{\bm{k}} \tilde{\chi}_{\bm{k}}$ with
\begin{equation}
\mathcal{A}_{\bm{k}} = -i\sigma_{z} U^{-1}_{\bm{k}} \frac{\partial U_{\bm{k}}}{\partial \bm{k}} 
\end{equation}
being the connection associated with the transformation $U_{\bm{k}}$.

Among the various representations, there is one specific basis, where the Hamiltonian in Eq.~(\ref{eq:Hmag}) becomes diagonal.  This basis is reached by choosing $\tanh 2\theta_{\bm{k}} = |B_{\bm{k}}|/A_{\bm{k}}$ and  $\phi_{\bm{k}} = \arg B_{\bm{k}}$, which gives
\begin{equation}
\hat{H} = \sum_{\bm{k}} \varepsilon_{\bm{k}} \left(\alpha^{\dag}_{\bm{k}} \alpha_{\bm{k}} +  \beta^{\dag}_{-\bm{k}} \beta_{-\bm{k}} \right),
\end{equation}
where $\varepsilon_{\bm{k}} = \sqrt{A_{\bm{k}}^{2} - |B_{\bm{k}}|^{2}}$ is the magnon energy dispersion.  To find the expression for the spin current in this basis, we notice that in Eq.~(\ref{eq:Jsmt})
\begin{equation}
-\frac{\partial \hat{\bm{A}}}{\partial t} = i[\hat{\bm{A}}, \hat{H}] = 
\sum_{\bm{k}} (\alpha_{\bm{k}}^{\dag}, \beta_{-\bm{k}})
\left(
\begin{array}{cc}
0 & \bm{K}^{*}_{\bm{k}} \\
\bm{K}_{\bm{k}} & 0
\end{array}
\right)
\left(
\begin{array}{c}
\alpha_{\bm{k}} \\
\beta_{-\bm{k}}^{\dag}
\end{array}
\right),
\end{equation}
is purely off-diagonal with $\bm{K}_{\bm{k}} = e^{i\phi_{\bm{k}}} \left[ \varepsilon_{\bm{k}}^{-1}(A_{\bm{k}} \partial_{\bm{k}} |B_{\bm{k}}| - |B_{\bm{k}}| \partial_{\bm{k}} A_{\bm{k}}) - i|B_{\bm{k}}| \partial_{\bm{k}} \phi_{\bm{k}} \right]$.  Therefore,  the total magnon spin current is written as 
\begin{equation}
\hat{\bm{J}}_{s} = 
\sum_{\bm{k}} (\alpha_{\bm{k}}^{\dag}, \beta_{-\bm{k}})
\left(
\begin{array}{cc}
\bm{v}_{\bm{k}} & \bm{K}^{*}_{\bm{k}} \\
\bm{K}_{\bm{k}} & \bm{v}_{\bm{k}}
\end{array}
\right)
\left(
\begin{array}{c}
\alpha_{\bm{k}} \\
\beta_{-\bm{k}}^{\dag}
\end{array}
\right),
\end{equation}
where $\bm{v}_{\bm{k}} = \partial_{\bm{k}} \varepsilon_{\bm{k}}$ is the group velocity of magnons \cite{Proskurin2018a}.  This expression generalizes two contributions to the spin current in Eqs.~(\ref{eq:J00}) and (\ref{eq:J10}) identified earlier in our semi-classical approach.

\subsubsection{Nonlinear response theory for magnon spin photocurrents}
By using semi-classical equations of motion in Sec.~\ref{sec:4.2}, we have already demonstrated that magnon spin photocurrent is the second order effect in the magnetic field of the electromagnetic wave.  Here, we show how the process of photo-excitation can be described via the nonlinear response theory.

Considering interaction of magnons with the electromagnetic wave as a perturbation, we can express the excited spin current using the second-order Kubo formula \cite{Tiablikov2013}
\begin{multline}
\label{kubo}
\langle \hat{\bm{J}}_{s}(t) \rangle = -\sum_{\omega_{1}\omega_{2}} \int_{-\infty}^{t} dt_{1} \int_{-\infty}^{t_{1}} dt_{2}
e^{\epsilon(t_{1} + t_{2} - t)}  e^{i\omega_{1}t_{1} + i\omega_{2}t_{2}} \\
\times \left\langle
\left[
\left[\hat{\bm{J}}_{s}(t), \hat{H}_{I}^{(\omega_{1})}(t_{1})\right], \hat{H}_{I}^{(\omega_{2})}(t_{2})
\right]
\right\rangle, \quad \epsilon \to 0^{+},
\end{multline}
where the interacting part of the Hamiltonian is taken in the form of dipole interaction between the magnetic field vector $\bm{\mathfrak{B}}_{\bm{k}}(\omega)$ and the local magnetization of the antiferromagnet, $\hat{H}_{I} = -g\mu_{B}\sum_{i} \bm{\mathfrak{B}}(t,\bm{r}_{i})(\bm{S}_{iA} + \bm{S}_{iB})$, where $g$ is the Land\'{e} factor.  In terms of magnon operators, it is expressed as
\begin{equation}
\hat{H}_{I}^{(\omega)} = -g\mu_{B}\sqrt{\frac{S}{2}}\sum_{\bm{k}} \left[
\mathfrak{B}_{\bm{k}}^{(-)}(\omega) \left(a_{\bm{k}} + b_{-\bm{k}}^{\dag}\right) + \mbox{H.c.}
\right].
\end{equation}
In Eq.~(\ref{kubo}), the operators are in the Heisenberg picture, e. g.  $\hat{H}_{I}^{(\omega_{1})}(t_{1}) = \exp(i\hat{H}t_{1}) \hat{H}_{I}^{(\omega_{1})} \exp(-i\hat{H}t_{1})$, and the statistical average is with the density matrix of the noninteracting system $\rho_{0} = \exp(-\hat{H}/k_{B}T)$.

Straightforward algebra shows that the spin current is calculated from Eq.~(\ref{kubo}) as follows \cite{Proskurin2018a}
\begin{multline}
\label{eq:Jss}
\langle \hat{\bm{J}}_{s}(t) \rangle =\frac{1}{4}\sum_{\omega\bm{k}}
\left[
\frac{\bm{v}_{\bm{k}} \mu_{\bm{k}}}{(\varepsilon_{\bm{k}} - \omega)^{2} + \Gamma^{2}} +
\frac{\bm{v}_{\bm{k}} \mu_{\bm{k}}}{(\varepsilon_{\bm{k}} + \omega)^{2} + \Gamma^{2}} 
\right.
\\
\left. +
\frac{\lambda_{\bm{k}} \bm{K}_{\bm{k}}}{(\varepsilon_{\bm{k}} - \omega - i\Gamma)(\varepsilon_{\bm{k}} + \omega - i\Gamma)} +
\frac{\lambda^{*}_{\bm{k}} \bm{K}^{*}_{\bm{k}}}{(\varepsilon_{\bm{k}} - \omega + i\Gamma)(\varepsilon_{\bm{k}} + \omega + i\Gamma)}
\right] \bm{h}_{\bm{k}}^{(-)}(\omega) \bm{h}_{-\bm{k}}^{(+)}(-\omega),
\end{multline}
where $\bm{h} = -g\mu_{B}\sqrt{2S}\bm{\mathfrak{B}}$, $h^{(\pm)} = h^{x} \pm h^{y}$, and the coefficient are given by
\begin{eqnarray}
\mu_{\bm{k}} &=& \frac{A_{\bm{k}} - |B_{\bm{k}}|\cos\phi_{\bm{k}}}{\sqrt{A_{\bm{k}}^{2} - |B_{\bm{k}}|^{2}}}, \\
\lambda_{\bm{k}} &=& e^{-i\phi_{\bm{k}}} \left( \frac{A_{\bm{k}}\cos\phi_{\bm{k}} - |B_{\bm{k}}|}{\sqrt{A_{\bm{k}}^{2} - |B_{\bm{k}}|^{2}}}
- i \sin\phi_{\bm{k}} \right).
\end{eqnarray}

This expression contains two kinds of terms.  The first is proportional to the group velocity of magnons, and, therefore, can be associated with actual motion of magnon wave packets.  The second, proportional to $\bm{K}_{\bm{k}}$, is related to intersublattice dynamics; it contains the phase gradient, $\partial_{\bm{k}} \phi_{\bm{k}}$.  This phase can be interpreted as an offset in dynamics of the magnetizations on $A$ and $B$ sublattices given by $a_{\bm{k}}(t) \sim \exp(i\varepsilon_{\bm{k}}t)$ and $b^{\dag}_{-\bm{k}}(t) \sim \exp(i\varepsilon_{\bm{k}}t - i\phi_{\bm{k}})$ respectively.  It may be accumulated as a result of the DMI combined with a specific lattice configuration \cite{Kawano2018}, or be generated by the external electric field via the Aharonov-Casher effect \cite{Nakata2017, Owerre2017}.

In the case when both $\bm{v}_{\bm{k}}$ and $\bm{K}_{\bm{k}}$ are odd under the transformation $\bm{k} \to -\bm{k}$, the spin current is determined by the asymmetric part of $\bm{h}_{\bm{k}}^{(-)}(\omega) \bm{h}_{-\bm{k}}^{(+)}(-\omega)$, which is proportional to $i[\bm{h}_{\bm{k}}^{*}(\omega) \times \bm{h}_{\bm{k}}(\omega)]_{z}$.  In the limiting case $\Gamma \to 0$ and $\phi_{\bm{k}} = 0$, we can combine both kinds of terms in Eq.~(\ref{eq:Jss}), which eventually gives
\begin{equation}
\label{eq:Jss1}
\langle \hat{\bm{J}}_{s}(t) \rangle = \frac{i}{2}\sum_{\omega\bm{k}} 
\frac{q_{\bm{k}}^{2} \partial_{\bm{k}} p_{\bm{k}} - \omega^{2} \partial_{\bm{k}}q_{\bm{k}}}{(\varepsilon_{\bm{k}}^{2} - \omega^{2})^{2}}
[\bm{h}_{\bm{k}}^{*}(\omega) \times \bm{h}_{\bm{k}}(\omega)]_{z},
\end{equation}
where $p_{\bm{k}} = A_{\bm{k}} + |B_{\bm{k}}|$ and $q_{\bm{k}} = A_{\bm{k}} - |B_{\bm{k}}|$, which coincides with Eq.~(\ref{eq:Js1}) obtained from the semi-classical equations of motion \cite{Proskurin2018a}.

\subsection{Magnon spin photocurrents in antiferromagnetic insulators and low dimensional materials}
We have demonstrated that in antiferromagnetic materials magnon spin currents contain intraband terms, proportional to the group velocity of magnons, and interband terms, which by analogy to the relativistic mechanics can be associated with the \emph{Zitterbewegung} effect of magnons.  The latter is proportional to the fast-oscillating factors, which makes these terms irrelevant as far as response to a static perturbation is concerned.  For the thermal excitation of spin currents, for example, the antiferromagnetic spin current can be taken in the form of Eq.~(\ref{eq:J00}) \cite{Rezende2016, Cheng2016a, Zyuzin2016}.

The response to a dynamic perturbation is different.  Since spin photocurrent is the second-order effect, the interband terms that oscillate at the double frequency should be taken into account together with the intraband contributions, so that the resulting response current is given by Eq.~(\ref{eq:Jss1}).

For practical applications, the most interesting frequency region is near the antiferromagnetic resonance, $\omega \approx \varepsilon_{\bm{k}}$. In this area, the response current is resonantly amplified and determined by the damping of the material.  In the case of ballistic magnon transport, when $\varepsilon_{\bm{k}} \gg \Gamma$, we can replace $\omega - \varepsilon_{\bm{k}} \pm i \Gamma \to \pm i\Gamma$ and $\omega + \varepsilon_{\bm{k}} \pm i \Gamma \to 2\omega_{r}$ near the resonance $\omega_{r}$.  In this limit, the dominant contribution in Eq.~(\ref{eq:Jss}) comes from the first term proportional to $\bm{v}_{\bm{k}}$
\begin{equation}\label{eq:Jres}
\langle \hat{J}_{s} \rangle_{\mathrm{res}} \approx \frac{iq_{\bm{k}}}{4\hbar\omega_{r}} \frac{v_{\bm{k}}}{\Gamma^{2}} 
[\bm{h}^{*}(\omega_{r}) \times \bm{h}(\omega_{r})]_{z},
\end{equation}
where we used monochromatic microwave field with $\bm{h}_{\bm{k}}(\omega)$ \cite{Proskurin2018a}.  This expression allows to estimate the order of magnitude for the spin photocurrent excited with circularly polarized light as $\langle \hat{J}_{s} \rangle_{\mathrm{res}} \approx \chi g^{2} \mu_{B}^{2} J_{1} S^{2} c_{s} I_{B}/(2a_{0}c^{2}\hbar\eta^{2}\omega_{r})$, where we take $\Gamma = \hbar \eta \omega_{r}$, $\chi = \pm$ denotes helicity of the wave, $I_{B} = |\bm{\mathfrak{B}}(\omega_{r})|^{2}$ is intensity, and linear magnon energy disperison is implied, $|v_{\bm{k}}| = c_{s}$.  For a typical material with $c_{s} = 3 \times 10^{4}$~m/s, $J_{s} = 200$~K, $\omega_{r} = 3 \times 10^{13}$~s$^{-1}$, $\eta = 10^{-4}$, and $a_{0} = 0.5$~nm, we estimate $\langle \hat{J}_{s} \rangle_{\mathrm{res}} \approx 1.5 \times 10^{4}$~A/m$^{2}$ (in electric units $e/\hbar$) for the microwave field strength $|\bm{\mathfrak{B}}| \approx 10$~mT.

Relative contributions of different terms in Eq.~(\ref{eq:Jss}) depend on the lattice configuration and on the details of microscopic interactions.   We may expect that in low-dimensional antiferromagnets interband contribution determined by the phase gradient becomes more significant.  We can separate this contribution from Eq.~(\ref{eq:Jss}) as follows
\begin{equation}
\label{eq:Jss2}
\langle \hat{\bm{J}}_{s} \rangle_{\phi} =\frac{1}{2}\sum_{\omega\bm{k}}
\frac{|B_{\bm{k}}| \sin \phi_{\bm{k}} \partial_{\bm{k}} \phi_{\bm{k}}}{\omega^{2} - \varepsilon_{\bm{k}}^{2}}
 \bm{h}_{\bm{k}}^{(-)}(\omega) \bm{h}_{-\bm{k}}^{(+)}(-\omega).
\end{equation}

\begin{figure}[t]
	\sidecaption[t]
	\includegraphics[scale=0.35]{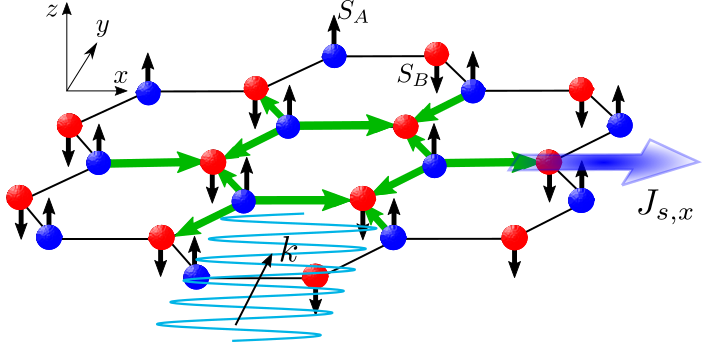}
	\caption{Two-dimensional antiferromagnetic insulator with two magnetic 
		sublattices $\bm{S}_{A}$ and $\bm{S}_{B}$ on the honeycomb lattice.  Green arrows show the DMI configuration.  The sign of $D_{ij}$ is positive for 
		$i \to j$ pointing from $A$ to $B$.}
	\label{fig:4}
\end{figure}

Let us find a model system where this term in the spin current can be excited individually. 
For this purpose, we consider a two-dimensional antiferromagnet  on the honeycomb lattice, as schematically shown in Fig.~\ref{fig:4}.  This model is interesting because antiferromagnetic magnons on the honeycomb lattice have finite $\phi_{\bm{k}}$ even without DMI. Indeed, straightforward algebra shows that $B_{\bm{k}} = J_{1}SC_{\bm{k}}$, where the structure factor is $C_{\bm{k}} = 2\cos(k_{x}/2) \cos(\sqrt{3}k_{y}/2) -1 + 2i \sin(k_{x}/2)[\cos(k_{x}/2) - \cos(\sqrt{3}k_{y}/2)]$, which in the long-wavelength limit gives the phase $\phi_{\bm{k}} \approx k_{x}(3k_{y}^{2} - k_{x}^{2})/8$.

Note that $\phi_{\bm{k}}$ is odd under $\bm{k} \to -\bm{k}$. In order to break this symmetry, we add the specific DMI configuration $D_{ij}(\bm{S}_{i} \times \bm{S}_{j})_{z}$ between the nearest neighboring sites $i$ and $j$ on the honeycomb lattice, such as $D_{ij} = D$ if  $i \in A$ and $j \in B$, and $D_{ij} = -D$ otherwise. Adding such term does not modify the energy dispersion, but instead leads to the constant phase accumulation $B_{\bm{k}} = J_{1}SC_{\bm{k}}\exp(i\phi_{0})$ where $\tan \phi_{0} = D/J_{1}$. In this case, $\sin(\phi_{\bm{k}} + \phi_{0}) \partial_{k_{x}} \phi_{\bm{k}}$ remains finite even in the $k_{x} \to 0$ limit. Therefore, by using Eq.~(\ref{eq:Jss2}), we are able to excite magnon spin current along $x$ by the linearly polarized electromagnetic wave propagating along the  $y$ axis, see Fig.~\ref{fig:4}. The magnitude of the spin current is estimated as $\langle \bm{J}^{x}_{s} \rangle_{\phi} \approx 3g^{2}\mu_{B}^{2}J_{1}S/(8\hbar^{2}c^{2})\sin \phi_{0} I_{B} \omega^{2}/(\omega^{2} - \varepsilon_{\bm{k}}^{2})$, and its sign is proportional to the sign of $\phi_{0}$.

\section{Conclusions}
\label{sec:5}
We have discussed how symmetry analysis can help to bring new ideas from optics to antiferromagnetic spintronics.  Our discussion started with an observation that a formal similarity between the electromagnetic field and spin waves in an antiferromagnetic insulator allows to find a generalization of optical chirality.  This forms a background for establishing a link between optics of chiral metamaterials and magnonics.  For example, spin wave absorption in chiral antiferromagnets can be described in the same terms as the electromagnetic energy dissipation in metamaterials.  Moreover, in antiferromagnets a pure spin current can provide a chiral symmetry breaking in a controllable way through the spin torque mechanism.

Fundamentally, this follows from the fact that spin currents are truly chiral; they have the same $\mathcal{P}\mathcal{T}$ transformation properties as e.g. optical chirality density. 
The latter suggests that chiral electromagnetic fields can be used for magnon spin current generation. We discussed that a direct magnon spin current appears as a second-order response to the circularly polarized microwave field, which frequency is near the antiferromagnetic resonance.  The direction of the current is determined by helicity of light that makes it similar to the circular photogalvanic effect in metals.

Lastly, we discuss how magnon spin currents in antiferromagnets have an interesting dynamics that can come into play for photo-excitation. Besides the transport terms proportional to the group velocity of the spin waves, there is a contribution from the trembling  motion of magnons, which can be identified by analogy with motion of ultra-relativistic particles. Although these fast oscillating terms can be safely omitted in some applications, they contribute to the photo-excitation process.

\begin{acknowledgement}
  R.L.S. acknowledges the support of the Natural Sciences and
  Engineering Research Council of Canada (NSERC) RGPIN 05011-18.
\end{acknowledgement}

\section*{Appendix: Magnon spin current definition from the
  antiferromagnetic Lagrangian}
\addcontentsline{toc}{section}{Appendix} Let us consider a classical
spin model for an antiferromagnet with two sublattices $\bm{S}_{A}$
and $\bm{S}_{B}$ with the energy given by
\begin{equation}
  \mathcal{H}_{\mathrm{AFM}} = J\sum_{\langle ij \rangle}\bm{S}_{i} \cdot \bm{S}_{j} - K \sum_{i}(S^{z}_{i})^{2},
\end{equation}
where $J > 0$ is a nearest neighboring exchange interaction, $K$ is
the anisotropy constant along the $z$-axis, and summation is over the
nearest neighboring sites on the $A$ and $B$ sublattices. For
simplicity of notations, we consider one-dimensional arrangement of
$\bm{S}_{i}$ along $x$. Semi-classical dynamics of this model can be
captured from the following Lagrangian \cite{Tvetev2016}
\begin{multline}
  \mathcal{L} = \int dx \left[ \rho \bm{M} \cdot \left(\bm{L} \times
      \frac{\partial \bm{L}}{\partial t}\right)
    -\frac{a}{2}|\bm{M}|^{2} \right.
  \\ 
  \left.  - A \frac{\partial }{\partial x} \left[\left(\frac{\partial
          \bm{L}}{\partial x}\right)^{2} - \left(\frac{\partial
          \bm{M}}{\partial x}\right)^{2}\right] - \ell \bm{M} \cdot
    \frac{\partial \bm{L}}{\partial x} +
    \frac{\tilde{\beta}}{2}(\bm{L}^{z})^{2} \right],
\end{multline}
where $\bm{M} = \frac{1}{2S}(\bm{S}_{A} + \bm{S}_{B})$ and
$\bm{L} = \frac{1}{2S}(\bm{S}_{A} - \bm{S}_{B})$, which satisfy the
constraints $\bm{M} \cdot \bm{L} = 0$ and
$\bm{M}^{2} + \bm{L}^{2} = 1$. The parameters of the Lagrangian are as
follows: $\rho = 2\hbar S$, $a = 8JS^{2}$, $\ell = 2JS^{2}a_{0}$,
$A = JS^{2}a_{0}^{2}$, and $\beta = 4KS^{2}$. Note that this
expression contains so-called topological term proportional to
$\ell$, which breaks the inversion symmetry in the Lagrangian
\cite{Tvetev2016}.

The expression for the spin current can be obtained applying the
Noether's theorem to the Lagrangian transformation under the local
infinitesimal rotation around $z$
\begin{eqnarray}
  \bm{M} &\to& \bm{M} + \delta \phi (\hat{\bm{z}} \times \bm{M}), \\
  \bm{L} &\to& \bm{L} + \delta \phi (\hat{\bm{z}} \times \bm{L}),
\end{eqnarray}
where $\delta \phi(x)$ is the local rotation angle.  The corresponding
change in the Lagrangian is given by
\begin{multline}
  \delta \mathcal{L} = - \int dx \delta \phi \left \lbrace \rho
    \frac{\partial}{\partial t} \left[ M^{z}(1 - |\bm{M}|^{2})\right]
  \right.
  \\ 
  \left.  - A\frac{\partial }{\partial x} \left[(\hat{\bm{z}} \times
      \bm{L}) \cdot \frac{\partial \bm{L}}{\partial x} - (\hat{\bm{z}}
      \times \bm{M}) \cdot \frac{\partial \bm{M}}{\partial x} \right]
    - \ell \frac{\partial }{\partial x} \left[\bm{M}\cdot(\hat{\bm{z}}
      \times \bm{L})\right] \right\rbrace,
\end{multline}
which gives the following expression for the spin current density
\begin{equation}
  j_{s}^{z} = -A \hat{\bm{z}} \left[\left(\bm{L} \times \frac{\partial \bm{L}}{\partial x}\right)
    - \left(\bm{M} \times \frac{\partial \bm{M}}{\partial x}\right) \right]
  - \ell \hat{\bm{z}} \cdot (\bm{L} \times \bm{M}).
\end{equation}
The first term in this expression is consistent with the expression
for the spin current obtained from the equations of motion. The second
is the contribution from the topological terms, which has different
symmetry. In particular, it changes the sign if we interchange
$\bm{S}_{A}$ and $\bm{S}_{B}$.

\bibliographystyle{spphys} \bibliography{chapter}

\end{document}